\documentclass{amsart}
\usepackage[scaled]{helvet}
\usepackage[utf8]{inputenc}

\usepackage{amsmath}
\usepackage{amsthm}
\usepackage{setspace}
\usepackage{appendix}
\usepackage{ifthen}
\usepackage{amssymb}
\usepackage{comment}
\usepackage{float}
\usepackage{url}
\PassOptionsToPackage{hyphens}{url}
\numberwithin{equation}{section}
\usepackage{pdfpages}
\DeclareMathOperator\erf{erf}

\newtheoremstyle{break}
  {\topsep}{\topsep}%
  {\upshape}{}%
  {\bfseries}{}%
  {\newline}{}%
\theoremstyle{break}

\newtheorem{theorem}{Theorem}[section]
\newtheorem{lemma}[theorem]{Lemma}
\newtheorem*{lemma*}{Lemma}

\newtheorem{coro}[theorem]{Corollary}
\newtheorem*{coro*}{Corollary}
\newtheorem{define}[theorem]{Definition}

\usepackage[a4paper, total={6in, 8in}]{geometry}
\begin{document}
\title{Stochastic Models for Replication Origin Spacings in Eukaryotic DNA Replication}
\author{Huw Day and Nina C. Snaith}
\begin{abstract}
    Replication of genetic material is an important process for all living organisms.  Origins of replication initiate the copying of DNA at many points on a chromosome, and it is the distribution of these points that is relevant here, as it presents us with an interesting stochastic process.  It was observed by Newman et al. \cite{Newman} that for various types of yeast cells, there were fewer very small inter-origin spacings, and fewer very large inter-origin spacings in the replication origin data  than would be expected if the origins were uncorrelated, random points.  We propose a very simple stochastic model for DNA replication and determine that this probabilistic process produces replication origins that display repulsion between origins and relative scarcity of large spacings. We detail some connections between this model and existing polynuclear or polymer growth models. 
\end{abstract}

\maketitle

\section{Introduction to Eukaryotic DNA Replication}
\label{sect:intro}

Cells are constantly replicating whenever organisms need to grow, repair or reproduce. Whenever a cell replicates, it has to copy the genetic information in the cell, which is stored as DNA. How it copies this efficiently and successfully using origins of replication scattered along chromosomes is a widely studied problem in genetics (for an overview, see  Chapter 6 of \cite{lodish2008molecular} or section H of \cite{irving2004bios}, and for a review of recent literature see \cite{kn:hu_stillman}). 

DNA is the genetic blueprint which allows for the creation of healthy and functioning cells. It contains all the information necessary for cells to perform various functions that are essential for life. DNA is stored as chromosomes. In eukaryotes (this is most types of organisms, including plants, fungi and animals), these chromosomes are linear in structure. In contrast, prokaryotes (bacteria and some single celled organisms) typically store DNA in circular chromosomes. DNA contains genes which provide the information for individual building blocks of the cell.

If we consider the cell as a factory (a producer of goods for the organism) and genes are the instructions about how to carry out various procedures, chromosomes are the instruction booklets. The genome is the whole set of DNA instructions found in a cell. DNA replication is a crucial part of cell replication, which occurs in growth and repair of any organism. 

\begin{figure}[H]
    \centering
    \includegraphics[scale=0.45]{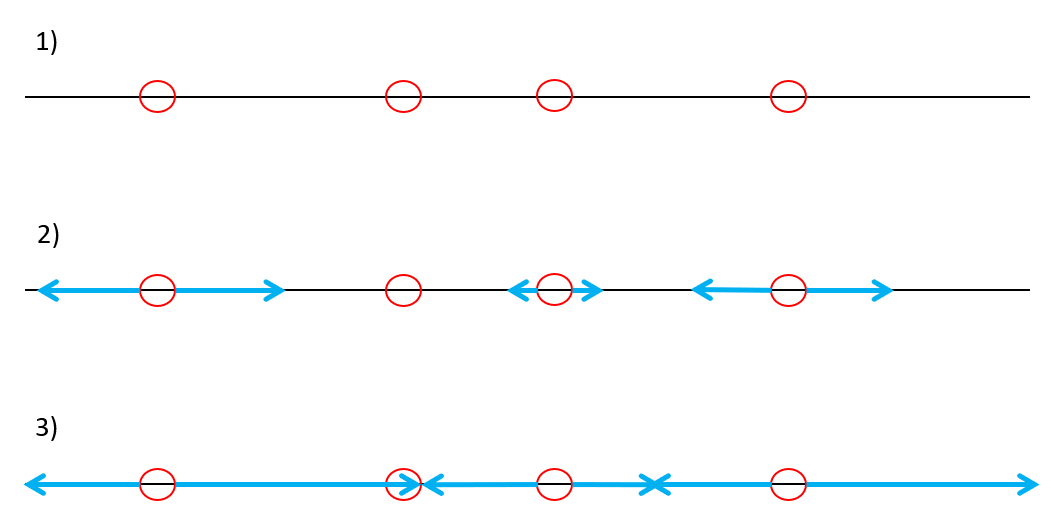}
    \caption{A schematic representation of the system we will attempt to model. This is inevitably an oversimplification of a part of the entire process of DNA replication. The basis of this description is taken from \cite{Newman}.
    Chromosomes in eukaryotes can be well modelled by a line and replication origins by small intervals on the line (in our calculations we model them as single points, but for the purpose of clarity in the diagram we draw red circles). In the first diagram we see licensed replication origins spaced out along the chromosome.
    Once replication begins, replication origins trigger at various times, sending a bidirectional replication fork which reads outwards along the chromosome. Thus the blue line is replicated DNA and the black is not. In the second diagram, all but the second origin have triggered.
    Eventually, the replication forks `read' the entire chromosome, represented by the entirety of the black line being covered by blue arrows in the third diagram. Note also that the second origin failed to trigger before replication forks from neighbouring origins reached it before it triggered. For our mathematical model, we will focus on the origins that trigger before they are read over, and we will name such points `active' points.}
    \label{fig:DNADiagram}
\end{figure}

 Replication is a hugely complicated process, the details of which vary widely from organism to organism, as can be seen by the vast literature on the subject.  We will consider a very simplistic view, looking at the most fundamental aspects that are common across most organisms.  On a linear chromosome, instead of going top to bottom, copying initiates at multiple points along the chromosome. These points are called replication origins. In each replication cycle, only some of these replication origins are licensed to trigger but some are not. When replication begins, the licensed origins will each trigger after some point in time. When an origin triggers, it sends out a replication fork, spreading out in both directions along the chromosome. This replication fork  reads and duplicates the genetic information it passes over. If a replication fork passes over a licensed replication origin before it triggers, we say the origin will become inactive (or passive) and will not trigger in this cycle of replication. We call licensed origins which trigger before they are made inactive, `active' origins. Our interest will solely be on how these active origins are distributed, as seen in Figure \ref{fig:DNADiagram}.

An issue with replication forks however, is that they are unreliable. This means there is a chance that either end of the fork may stall irreversibly in place. If one replication fork stalls, this is not a major problem, since a replication fork approaching it from the other direction will still replicate the rest of the genetic information between them. The issue comes when two replication forks replicating each other both stall before colliding, leaving a section of unread genetic information in between them. We refer to these as double stall events.

If a double stall occurs when reading a chromosome, then the replicated DNA will be incomplete \cite{Newman}. This can lead to mutations and in some cases genetic diseases. There are quality checkpoints and repair mechanisms in place to mitigate the damage from double stall events.   It is common in the genetics literature, such as  \cite{Newman}, \cite{Overcook}, to model these systems as discrete in both time and space. Both consider the problem of double stall events and how these impede both successful and timely DNA replication. It is also important to note that double stall events are not the only impediment to successful replication.

Due to natural selection, we might expect that the distribution of replication origins on chromosomes is as efficient as it could be given certain external constraints. It is an interesting mathematical problem to study the statistics of these replication origins in various organisms. Additionally, better understanding these statistics could provide a better understanding for creating new treatments to treat genetic diseases \cite{genetherapy}.

\begin{figure}
    \centering
    \includegraphics{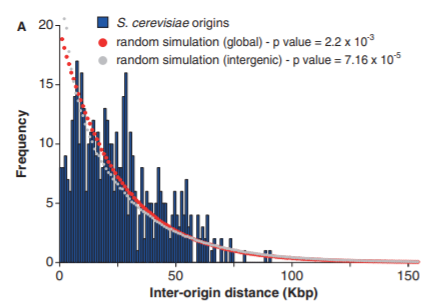}
    \caption{This Figure is Figure 3A from \cite{Newman}. Original caption: ``Inter-origin spacings in the S. cerevisiae genome. (A) Interorigin spacings in S. cerevisiae were calculated and assigned to different 1 kb bins. The frequency of origins in each bin is shown. Red dots: mean origin separation in a computer simulation where the same number of origins were placed at random on the whole S. cerevisiae genome. Grey dots: mean origin separation in a computer simulation where the same number of origins were placed at random only in the intergenic regions of the S. cerevisiae genome"}
    \label{fig:NewmanFig}
\end{figure}

There is a figure in \cite{Newman}, reproduced here in our Figure \ref{fig:NewmanFig},  which beautifully represents the distribution of spaces between replication origins in the yeast S. Cerevisiae. On the same plot is shown the distribution of spacings between random, uncorrelated points scattered through the whole genome (red), or just in the gaps between genes (grey). The clear gap between the histogram and the red and grey curves for small spacing size is evidence of repulsion between replication origins - meaning small spacings are less common than if the points were uncorrelated. From a biological perspective it seems inefficient for replication origins to be very close together, as the replication forks would almost immediately merge.  Data is very limited, but the figure makes it plausible that there are fewer large spacings than the uncorrelated points, and this is corroborated further elsewhere in \cite{Newman}.  The discussion of double stall events suggests that having large spacings between origins might be risky.  The phenomenon of a paucity of very small or very large spacings is commonly observed in the statistics of eigenvalues of random matrices, and this connection is explored in the companion paper \cite{kn:DaySnaith}. 

The question was raised in \cite{herrick2002kinetic} as to how such correlated replication origins, as seen in Figure \ref{fig:NewmanFig}, might arise in a biological system. In the current paper, we propose a very simple stochastic process to mimic the most basic attributes of DNA replication and aim to show that if even if the input to this process is completely uncorrelated points from a 2D Poisson process, the active origins where replication commences will be correlated in such away that they show fewer very small and fewer very large spacings. 

\subsection{Kolmogorov-Johnson-Mehl-Avrami (KJMA) Model}

There exist models for the nucleation and spread of various processes, for example, consider a tray of water suddenly reduced to subzero temperature at time $t=0$. A natural question to consider is what fraction of the water is frozen at time $t \geq 0$? This is reviewed in \cite{DNANuc1}.

The general model of this sort of problem, called Kolmogorov-Johnson-Mehl-Avrami (KJMA) Model or sometimes just Avrami Model has been studied in detail in the mid-1900s \cite{kolmo1937,william1939,avrami1941} with various applications, primarily in material sciences \cite{christian2002} but also in study of crystallisation structures \cite{yang1988}. We will outline this model using material from \cite{DNANuc1} and then see its applications to the DNA replication model as explained in \cite{DNANuc2}.

\begin{figure}[H]
    \centering
    \includegraphics[scale=0.5]{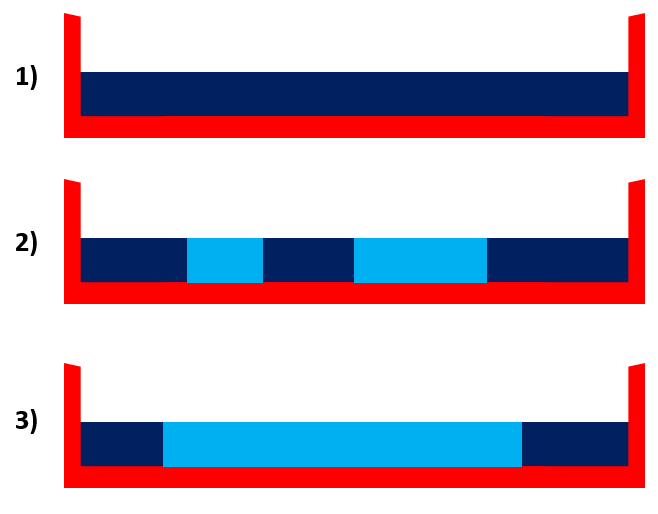}
    \caption{A (red) tray of (dark blue) water placed in sub zero temperature will begin freezing in different places at different times. Freezing (denoted with a lighter blue) will spread outwards overtime until all of the water is frozen. We see just water in 1), two points of nucleation/freezing occurring after some length of time in 2) and in 3) we see coalescence where two ice fronts meet after more time has passed.}
    \label{fig:ice}
\end{figure}

The three main mechanisms in the KJMA model are:
\begin{enumerate}
    \item Nucleation: where `freezing' begins at certain points on the water.
    \item Growth: where this freezing effect spreads from nucleation points in a bi-directional manner, creating an expanding island of ice.
    \item Coalescence: when two expanding nucleation fronts converge, growth terminates at their meeting point. Essentially when two bits of ice meet and form one big island out of two smaller islands.
\end{enumerate}
These are illustrated in Figure \ref{fig:ice}.

Nucleation is described with a point process of some form. This process is typically two dimensional with one dimension describing where freezing starts in blocks of ice and the other describing when the freezing starts at that time. Simpler models rely on Poisson point processes of uniform (i.e. homogeneous) intensity so that there is equal probability of nucleation for all locations. Some applications use inhomogeneous intensity, where the intensity of points depends on their locations and sometimes their time. In some stochastic models the points of coalescence also form a point process. However, it is not necessary to describe the coalescence point process to define the system.

Once a nucleation point has formed, it grows out in both directions with velocity $v$. Simpler applications have this velocity as constant and due to the geometry involved, it can often be useful to take this value to be $v=1$. More complicated applications might dictate that this velocity be inhomogeneous. This could easily correspond to an inhomogeneous intensity of nucleation points, for example if the water in our tray was at warmer temperatures in some areas along the tray, we would expect nucleation (i.e. freezing) to be less likely to occur in these warmer areas of the water and in the same manner we would expect the freezing growth to occur more slowly through these regions.

\begin{figure}[H]
    \centering
    \includegraphics{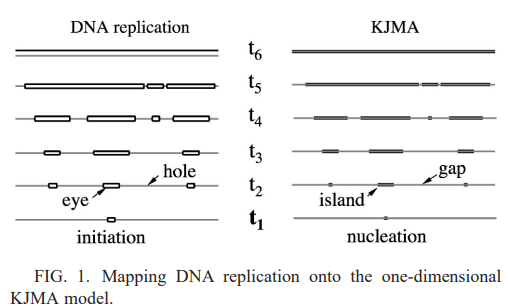}
    \caption{Figure taken from \cite{DNANuc2}. Original caption ``Mapping DNA replication onto the one-dimensional KJMA model." We see a clear correspondence between the nucleation/freezing in KJMA and the initiation/firing of replication origins in DNA replication. In the left side above, the ``eye", where the line is doubled indicates where DNA has already replicated. }
    \label{fig:DNAKJMA}
\end{figure}

As demonstrated in Figure \ref{fig:DNAKJMA}, the work in \cite{DNANuc2} makes a clear link between DNA replication and KJMA. Nucleation events are just like replication origins that fire, with this point process considered as a 2D Poisson point process of intensity $I$. The growth mechanism corresponds to the replication forks which spread out in a bi-directional manner. Coalescence is when two replication forks meet and the corresponding areas of the chromosome are read. Experimentally the replication fork speed $v$ can be treated as constant (see \cite{SpatioTemp} for more on this).

The method in \cite{DNANuc1} and \cite{DNANuc2} is to construct and solve a partial differential equation describing the evolution of quantities of interest, such as the fraction of the chromosome that has been replicated at a given time and the density of stretches of unreplicated chromosome of a given length. 

 In addition, the authors of \cite{herrick2002kinetic} collect large amounts of data from incidences of DNA replication to make comparisons with a general KJMA model. They use these comparisons to attempt to calculate parameters of the KJMA model such as the island/replication fork velocity $v$ and the intensity of the nucleation events,  sometimes even assuming that the intensity is both spatially and temporally inhomogeneous.

In contrast, in this paper we introduce a similar model, but use the geometry of the system and probabilistic methods to explore the distribution of spacings between active origins where replication commences, to see if it broadly captures the key features of replication origin distribution.

\section{2D Poisson Process Exclusion Model}
\label{sect:model}

In this section we consider a stochastic model that seeks to be an analogue of the DNA replication process we are modelling. The point is to discover what characteristics of the statistical distribution of replication origins are captured by a relatively simple probabilistic model - particularly focusing on repulsion between neighbouring origins, and the large-spacing tail of the distribution.  

We consider random points in two dimensions (representing the replication origins) where the horizontal axis represents position on the chromosome and the vertical axis the time when each origin triggers and starts replication.   Using these two pieces of information for each point, we perform a non-uniform exclusion process to select so-called `active' points (those which trigger before they are made inactive by a replication fork approaching from a nearby origin) and exclude all other points 
 - the `passive' points. Unlike Newman et al.'s work \cite{Newman} we do not consider the role that stalling has on our model.

The most natural model would be   to fix the length of our chromosome to be some parameter $L>0$. The origins would all be located in $[0, L]$, and so the horizontal coordinate of our points would also lie in this range.  Whilst there are clearly limitations on the trigger times (the lifetime of a cell, for example), for our calculations, we can say that the trigger times could theoretically be anywhere in $[0, \infty)$, so the vertical axis is semi-infinite.

We imagine generating a 2D Poisson point process of parameter $1$ on this space \([0, L] \times [0, \infty)\). This means that given a suitably nice set \(A \subset [0, L] \times [0, \infty)\) with measure (i.e. Euclidean area) $|A|$, we understand that:

\begin{equation}
N_{A}:=\text{Number of Poisson points in A} \sim Poisson(|A|).
\end{equation}

Explicitly, the probability of there being $k$ points in the box $A$ with area/measure $|A|$ is described as:
\begin{equation}
\label{Prob}
P(N_{A}=k)=\frac{|A|^{k}e^{-|A|}}{k!} \text{ for } k=0, 1, 2 \ldots
\end{equation}

Once we have a number of points in $A$, these points are scattered randomly but uniformly - they are no more likely to be in one part of $A$ than any other. Furthermore, we have that given two disjoint subsets of our domain $A$ and $B$ (that is, two areas that do not overlap), the random variables \(N_{A}\) and \(N_{B}\) are independent. This uniquely defines our Poisson point process.

We end up with a collection of points in our domain. Now we seek to identify the so-called `active' points and eliminate the non-active/passive points that get replicated by a replication fork emanating from a neighbouring origin.

\subsection{Distinguishing Active and Passive Points}

Denote a given Poisson point \((x_{i}, t_{i})\) where the first entry indicates its position on the chromosome and the second indicates its trigger time.

\begin{figure}[H]
	\centering
		\includegraphics[scale=0.6]{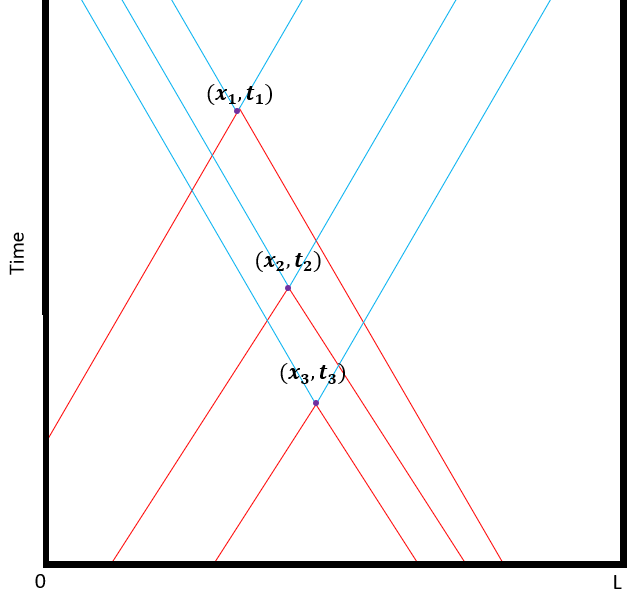}
	\caption{A basic realisation of the 2D Poisson process with $3$ points (marked by purple dots). Backward light cones are marked in red, forward light cones are marked in blue.
	In this instance we see that the highest point \((x_{1}, t_{1})\) contains the other $2$ points in its backwards cone, which means it would be made passive by \((x_{2}, t_{2})\) without the presence of \((x_{3}, t_{3})\). Similarly, \((x_{2}, t_{2})\) is made passive by the lowest point. We can say that the lowest point \((x_{3}, t_{3})\) is active and in this case makes the other $2$ points passive. Note that the angle at the top of a red triangle would be 90 degrees if we assume that the replication fork proceeds with speed 1.}
	\label{fig:GeomModel}
\end{figure}

We assume the speed of the replication fork emerging from each replication origin is constant and the same for all replication origins. (This is not an unreasonable assumption, see \cite{SpatioTemp} for more on this. Essentially, the average replication fork speed is found to not vary significantly). We will take this value to be $1$ for simplicity. The blue lines in Figures \ref{fig:GeomModel} and \ref{fig:Gap2} represent the progress of a replication fork along the chromosome over time.  Note that in these figures, the slope of these lines is not necessarily depicted as 1.

We take the definition of Forward and Backward Light Cones from \cite{PNG}. It is based on vocabulary from special relativity which will prove useful in describing aspects of our model:

\begin{define}[Forward and Backward Light Cones \cite{PNG}]

The forward light cone of a point $(x, t)$ is the set of points \(\{(x', t'): |x-x'|\leq t'-t\}\). The forward light cones track the progress of replication forks through space and time.
The backward light cone of a point $(x, t)$ is the set of points \(\{(x', t'): |x-x'|\leq t-t'\}\).

\label{def:LightCones}
\end{define}

If we consider this problem geometrically, we see in Figure \ref{fig:Gap2} that the isosceles right angled triangle underneath each origin \((x_{i}, t_{i})\) formed by two lines emitting from the point, of slope $+1$ and $-1$ respectively delineate its backward facing light cone as described in \cite{PNG}). This triangle will have vertices \((x_{i}-t_{i}, 0)\), \((x_{i}+t_{i}, 0)\) and \((x_{i}, t_{i})\). Upon closer inspection, it becomes apparent that any origin located in this triangle will make the point at the top of the triangle passive. The point \((x_{i}, t_{i})\) is active if and only if its backward facing light cone is empty of points.

Additionally, we see that if we draw the upward triangle from each point  \((x_{i}, t_{i})\), marked by lines of slope 1 and -1, we obtain its forward light cone. The forward light cone marks the progress of the replication forks through space and time. We see that any point in this forward cone is made passive by our original point. Since a passive point is always in the forward light cone of at least one active point, the lowest point in a system (i.e. the point with the smallest trigger time) is always active.

Our analysis of this model involves calculating the probability that backward light cones of Poisson points are empty of other points. This is complicated at the edges of the system when the backward light cone is not completely contained in the domain of the Poisson point process (see, for example, point $(x_1,t_1)$ in Figure \ref{fig:GeomModel}). We consider two ways to avoid this. The first, in Section \ref{sect:borderlessExc}, is to let $L$ go to infinity, which we call the borderless exclusion model.  The second, inspired by probabilistic models with similar features, is instead to choose a triangular domain that follows the lines of the backwards light cones, as we will see in Section \ref{sect:triangle}.

\subsection{Nearest Neighbour Spacing Calculation for the Borderless Exclusion Model} \label{sect:borderlessExc}

In this section we impose our exclusion model in a setting where we let the length of the system go to infinity. This has the drawback of not honouring the DNA replication setting as strongly, but  the nearest neighbour spacing has a closed form.

\subsubsection{Deriving the Nearest Neighbour Spacing Density}

We will now look at the distribution of spacings between neighbouring active points.  We will start by considering two points \((x_{1}, t_{1})\) and \((x_{2}, t_{2})\) that are both active and have no active points between them. Without loss generality say that \(x_{1}\leq x_{2}\).

\begin{figure}[H]
    \centering
    \includegraphics[scale=0.9]{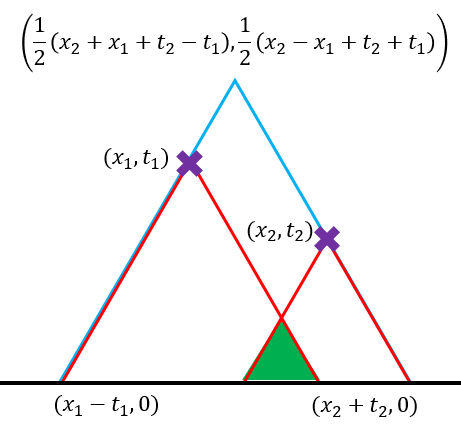}
    \caption{Two active points illustrated with purple crosses, their backward light cones in red and the meeting of their forward light cones in blue. The entire triangular shape must be empty in order for these two points to be neighbouring active points.}
    \label{fig:Gap2}
\end{figure}

Consider the area we require to be empty in order for this to be true. Looking at Figure \ref{fig:Gap2}, we definitely need the red backward light cones underneath our two points to be empty of points in order for them to be active.

In order to have no active points between these two points, we require that there are no points inside the rectangle shown in Figure \ref{fig:Gap2} with two red sides and two blue sides and our two active points on opposite corners. It is a rectangle because the the peak of both backward light cones are right angles if the replication fork speed is 1.

So in total, the entire triangle depicted in Figure \ref{fig:Gap2} must be empty of points in order for $(x_1,t_1)$ and $(x_2,t_2)$ to be neighbouring active points.  The width of this triangle is $x_{2}+t_{2}-(x_{1}-t_{1})$ and its height is half of this, so we have that the total area we require to be empty is:

\begin{equation}Area(x_1,x_2, t_{1}, t_{2}):=\frac{1}{4}(x_2-x_1+t_{1}+t_{2})^2.\end{equation}

We know for a 2D Poisson process of intensity $1$, the number of Poisson points in an area  of size $|A|$ is Poisson random variable with parameter $|A|$ or equivalently, for a non-negative integer $k$ we have the probability:

\begin{equation}
    {\rm P}(\text{Number of Poisson points in box of area } |A|  =k)=\frac{|A|^{k}e^{-|A|}}{k!}.
\end{equation}

The probability  of an area being empty ($k=0$) in a Poisson process of intensity $1$ is just $e^{-Area}$, so 

\begin{equation}
{\rm P}(Area(x_1,x_2,t_1,t_2)\; \text{empty of points})=\exp{\left(-\frac{1}{4}(x_2-x_1+t_{1}+t_{2})^2\right)}.
\label{eq:borderlessdensity}
\end{equation}

 Our goal is to find the probability density function of nearest neighbour spacings: that is, given an active point at some position, say $x_1$, we want to know how likely it is that the next active point is a distance $s$ away, irrelevant of the times at which they commence replication. To do this we begin by considering the probability that a Poisson point lands in the vicinity of $(x_1,t_1)$, another in the vicinity of $(x_2,t_2)$ and that they are neighbouring active points.

The probability of finding one or more points in a vanishingly small area $\delta$ is given by: $1-e^{-\delta}\approx \delta$. This is essentially the probability of finding one point in the area $\delta$, because the probability of finding more than one point is negligible. 



We are looking for one point in an area $dx_1dt_1$ at $(x_1,t_1)$ and another in $dx_2dt_2$ near $(x_2,t_2)$, and the probability that these points are neighbours and both active is captured by (\ref{eq:borderlessdensity}). Putting all of this together gives us the probability of an active point near $(x_1,t_1)$, another near $(x_2,t_2)$, and none in between:

\begin{equation}
    dx_{1}dt_{1}dx_{2}dt_{2}\exp{\left(-\frac{1}{4}(s+t_{1}+t_{2})^2\right)},
\end{equation}
where for convenience we denote the distance $x_2-x_1$ by $s$.

We now want to integrate over the allowed range for times $t_1$ and $t_2$.  We know that our times lie in \(0 < t_{1}, t_{2} <\infty\) but we also want both points to be active with respect to each other. This means that we want the height of the triangle that peaks at \(t_{1}\) to be lower than \(t_{2}\) when it coincides with \(x_{2}\), which means we want: \(t_{1}-s< t_{2}\). We also want the same to be true for \(t_{2}\), that is: \(t_{2}-s< t_{1}\). Thus given $t_1$,  the allowed range on $t_2$ is \(t_{1}-s < t_{2} < t_{1} +s\) but we note that we have to make sure \(t_{2} > 0\). Observe that $0$ is the dominant (i.e. larger) lower bound when \(t_{1}-s<0 \); that is, when \(t_{1}<s\).

So the probability of a Poisson point process having one active point in an in the interval $x_1$ to $x_1+dx_1$ and the next active point in the interval $x_2$ to $x_2+dx_2$, regardless of the times when replication initiates at these points, is



\begin{eqnarray}&&\Bigg(\int_{0}^{s} \int_{0}^{t_{1}+s} \exp\left(-\frac{1}{4}(s+t_{1}+t_{2})^2\right) dt_{2}dt_{1} \nonumber \\&&\qquad+ \int_{s}^{\infty}  \int_{t_{1}-s}^{t_{1}+s}  \exp\left(-\frac{1}{4}(s+t_{1}+t_{2})^2\right)dt_{2}dt_{1}\Bigg)dx_1 dx_2,
\label{eq:marginal}\end{eqnarray}
where still $s=x_2-x_1$.

It may seem that, in performing this integration, we are in danger of double-counting non-disjoint cases where two or more points fall in one of the intervals ($x_1$, $x_1+dx_1$) or ($x_2$, $x_2+dx_2$). These intervals represent an infinitely long vertical strip in the position-time plane, but we can instead consider the problem of a vertical strip with area that is small for small $dx_1$, say, but large height - essentially truncating the $t_1$ integral in (\ref{eq:marginal}).  The probability of finding active points very high in the $t$ direction is vanishingly small due to the size of the empty triangle this would require: the effect of truncating the integral is negligible thanks to the rapid decay of the integrand. The probability of finding more than one Poisson point in a strip of small area is negligible, so we don't need to worry about double-counting. 

In this model, the nearest neighbour spacing density of active points on a chromosome, $P(s)ds$, is defined as the probability that the next active point is between  $x_1+s$ and $x_1+s+ds$, given an active point at $x_1$.  We will denote by $\rho_{b}$ the density of active points on the chromosome; that is, the average number of active points per unit length on the $x$ axis.  The expected number of active points in the interval from $x_1$ to $x_1+dx_1$ is therefore $\rho_{b}dx_1$.  For an infinitesimal interval $dx_1$, the expected number of points is essentially the same as the probability of finding a point in that interval, due to the fact that there is a negligible probability of finding more than one point in the interval.  So, the probability of finding an active point in the interval from $x_1$ to $x_1+dx_1$ is $\rho_b dx_1$.  

Thus, to get  the nearest neighbour spacing distribution we calculate the required conditional probability by dividing (\ref{eq:marginal}) by $\rho_bdx_1$. 

\begin{eqnarray}P(s)ds&=&\frac{1}{\rho_b}\Bigg(\int_{0}^{s} \int_{0}^{t_{1}+s} \exp\left(-\frac{1}{4}(s+t_{1}+t_{2})^2\right) dt_{2}dt_{1} \nonumber \\&&+\int_{s}^{\infty}  \int_{t_{1}-s}^{t_{1}+s}  \exp\left(-\frac{1}{4}(s+t_{1}+t_{2})^2\right)dt_{2}dt_{1}\Bigg) ds,
\label{eq:Ps}\end{eqnarray}
where we have replaced $dx_2$ with $ds$.

Now we will calculate an exact expression for integrals in the outer bracket and then later we will use this to calculate the density $\rho_{b}$.

\begin{figure}[H]
           \centering
           \includegraphics[scale=0.4]{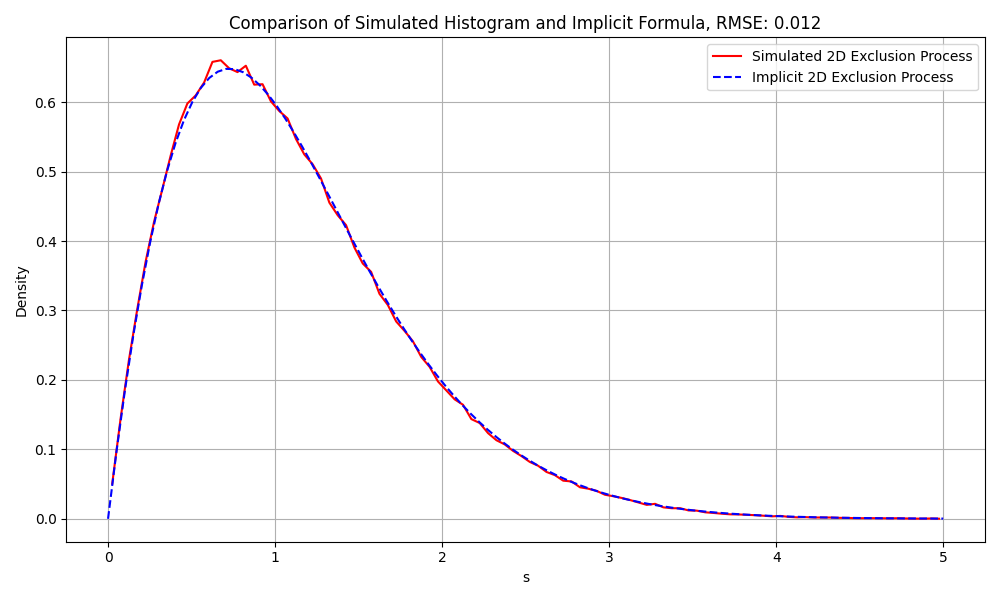}
            \caption{The blue curve is a numerical evaluation of the form of the nearest neighbour spacing, $P(s)$,    given in (\ref{eq:noboundary}). The red curve is the result of a numerical simulation of the exclusion model starting from a Poisson point process of unit intensity. }
            \label{fig:checkingnns}
        \end{figure}

\begin{theorem}[Nearest Neighbour Spacing for Active Points on a Borderless Exclusion Model]
\label{thm:NNSBorderless}
Consider a homogeneous 2D Poisson Point process of unit intensity imposed on $\mathbb{R}\times\mathbb{R}_{+}$. We apply the exclusion model described in Section \ref{sect:model}, defining active points to be those with  backwards light cone (with lines of slope $\pm 1$) empty of other Poisson points. $\rho_b$ is the spatial density of such active points. 

The nearest neighbour spacing, $P(s)ds$, is the probability that, given an active point, the next active point will be between distance $s$ and $s+ds$ away.  $P(s)$ is given by:

\begin{equation}
P(s)
\label{eq:noboundary}
    =\frac{1}{\rho_b}\left[s\sqrt{\pi}\left(1-2\erf{(s)}+\erf{\left(\frac{s}{2}\right)}\right)-2e^{-s^2}+2e^{-\frac{s^2}{4}}\right].
\end{equation}
The probability density given in (\ref{eq:noboundary}) is plotted numerically in Figure \ref{fig:checkingnns}.
\end{theorem}

\begin{proof}

We start from (\ref{eq:Ps}). Using the definition of the error function, 
\begin{equation}\erf{(z)}:=\frac{2}{\sqrt{\pi}}\int_{0}^{z}e^{-t^2}dt,\end{equation}
we notice that we can write the inner integrals in the form:
    \begin{equation}\label{eq:errorfunc}
        \int_{a}^{b}\exp{\left(-\frac{1}{4}(s+t_{1}+t_{2})^2\right)}dt_2
     =\sqrt{\pi} \left(\erf{\left(\frac{b+s+t_{1}}{2}\right)} -\erf{\left(\frac{a+s+t_{1}}{2}\right)}\right).   \end{equation}
   Thus,
    \begin{equation}
        \int_{0}^{t_{1}+s}\exp\left(-\frac{1}{4}(s+t_{1}+t_{2})^2\right)dt_{2}=\sqrt{\pi}\left(\erf{(s+t_{1})} -\erf{\left(\frac{s+t_{1}}{2}\right)}\right).
        \label{eq:1stintborderless}
    \end{equation}
    Next we integrate this expression with respect to $t_{1}$. 
    \begin{eqnarray}\label{eq:erf_int}
       \int_{0}^{s} \erf{\left(\frac{s+t_{1}}{c}\right)}dt_{1}&= &c \int_{\frac{s}{c}}^{\frac{2s}{c}} \erf{(x)}dx \\
       &=&cx \erf{(x)}\Bigg|_{\frac{s}{c}}^{\frac{2s}{c}} -\int_{\frac{s}{c}}^{\frac{2s}{c}} \frac{2 c x e^{-x^2}}{\sqrt{\pi}}dx\nonumber \\
       &=&cx \erf{(x)}\Bigg|_{\frac{s}{c}}^{\frac{2s}{c}}+\frac{c e^{-x^2}}{\sqrt{\pi}}\Bigg|_{\frac{s}{c}}^{\frac{2s}{c}}\nonumber \\
       &=&2s \erf{\left(\frac{2s}{c}\right)}-s \erf{\left(\frac{s}{c}\right)}+\frac{c}{\sqrt{\pi}}\left(\exp^{-\frac{4s^2}{c^2}}-e^{-\frac{s^2}{c^2}} \right),\nonumber
    \end{eqnarray}
    where $c$ can play the role of 1 or 2 in (\ref{eq:1stintborderless}).

    So
    \begin{eqnarray}\label{eq:term1}
      & \int_0^s\int_{0}^{t_{1}+s}&\exp\left(-\frac{1}{4}(s+t_{1}+t_{2})^2\right)dt_{2}ds= s \sqrt{\pi} \left(2 \erf{(2s)}- \erf{\left(s\right)} - 2 \erf{(s)}+\erf{\left(\frac{s}{2}\right)}\right)\nonumber\\
       &&\qquad
        +e^{-4s^2}-e^{-s^2}-2e^{-s^2}+2e^{-\frac{s^2}{4}}\\
       && =s \sqrt{\pi}\left(2 \erf{(2s)} - 3\erf{(s)}+\erf{\left(\frac{s}{2}\right)}\right)\nonumber \\
       &&\qquad+\exp{(-4s^2)}-3\exp{(-s^2)}+2\exp{\left(-\frac{s^2}{4}\right)}.\nonumber
    \end{eqnarray}
    Now to address the second integral in (\ref{eq:Ps}):
    \begin{equation}
        \int_{s}^{\infty} \int_{t_{1}-s}^{t_{1}+s}\exp{\left(-\frac{1}{4}(s+t_{1}+t_{2})^2\right)}dt_2dt_1 .
    \end{equation}
    Using (\ref{eq:errorfunc}),
    \begin{equation}
        \int_{t_{1}-s}^{t_{1}+s} \exp{\left(-\frac{1}{4}(s+t_{1}+t_{2})^2\right)}dt_2=\sqrt{\pi} \left(\erf{\left(s+t_{1}\right)} -\erf{\left(t_{1}\right)}\right).
        \label{eq:shiftederfs}
    \end{equation}
Performing the integral in $t_1$, we use integration by parts as before to obtain
    \begin{eqnarray}\label{eq:term2}
      &&  \sqrt{\pi}\int_{s}^{\infty} \left(\erf{\left(s+t_{1}\right)} -\erf{\left(t_{1}\right)}\right)dt_{1}=\nonumber\\&&\qquad\sqrt{\pi}\left[(t_{1}+s)\erf{(t_{1}+s)}+\frac{e^{-(t_{1}+s)^2}}{\sqrt{\pi}} - t_{1}\erf{(t_{2})}-\frac{e^{-t_{1}^2}}{\sqrt{\pi}} \right]_{s}^{\infty}\nonumber \\
      &&=\sqrt{\pi}s(1-2\erf{(2s)}+\erf{(s)})+e^{-s^2}-e^{-4s^2},
    \end{eqnarray}
    where we use the fact that the error function is defined so that $\erf{(\infty)}=1$. 

    Summing (\ref{eq:term1}) and (\ref{eq:term2}) in (\ref{eq:Ps}), we complete the proof. The resulting expression for $P(s)$ is computed numerically in Figure \ref{fig:checkingnns}.
\end{proof}

\subsubsection{Mean density of active points}
Using the preceding calculation of the nearest neighbour spacing, we calculate the mean density of points that are active. 

\begin{lemma}
    Given the nearest neighbour spacing, $P(s)$, from the borderless exclusion model, we can calculate the spatial density of active points $\rho_{b}$:
    \begin{equation}
    \rho_{b}=\frac{\sqrt{\pi}}{2}.
    \end{equation}
    \label{lem:rho}
\end{lemma}

\begin{proof}

We start with the result (\ref{eq:noboundary}),
\begin{equation}
P(s)
    =\frac{1}{\rho_b}\left[s\sqrt{\pi}\left(1-2\erf{(s)}+\erf{\left(\frac{s}{2}\right)}\right)-2e^{-s^2}+2e^{-\frac{s^2}{4}}\right].
\end{equation}
The nearest neighbour spacing is a probability density and so integrates to 1.  Thus we have that 
\begin{equation}
 \rho_b=\int_{0}^{\infty}s\sqrt{\pi}\left(1-2\erf{(s)}+\erf{\left(\frac{s}{2}\right)}\right)-2e^{-s^2}+2e^{-\frac{s^2}{4}}ds.
    \end{equation}


Using integration by parts twice:
\begin{equation}
    \int_0^{\infty} s\sqrt{\pi} \erf{(s/a)}ds=\frac{s^2}{2} \sqrt{\pi} \erf{(s/a)}\Big|_0^\infty-\frac{1}{a}\int_0^\infty s^2 e^{-s^2/a^2}ds,
\end{equation}
and
\begin{eqnarray}
    \frac{2}{a}\int_0^x s^2 e^{-s^2/a^2}ds&=&-ase^{-s^2/a^2}\Big|_0^x+a\int_0^x e^{-s^2/a^2}ds\nonumber \\
   &=& -ase^{-s^2/a^2}\Big|_0^x+a^2\int_0^{x/2} e^{-s^2}ds\nonumber \\
    &=&-ase^{-s^2/a^2}\Big|_0^x+a^2\frac{\sqrt{\pi}}{2}\erf({s/a})\Big|_0^{x}.
\end{eqnarray}
In addition
\begin{equation}
2\int_0^xe^{s^2/a^2}ds=a\sqrt{\pi}\erf(s/a)\Big|_0^x.
\end{equation}

So we can perform the integration to obtain



\begin{eqnarray}
  &&  \rho_b = \left[\frac{s^2 \sqrt{\pi}}{2}\left(1-2\erf{(s)}+\erf{\left(\frac{s}{2}\right)} \right)+\sqrt{\pi}\left(-\frac{\erf{(s)}}{2}+\erf{\left(\frac{s}{2}\right)}\right)\right.
\nonumber \\
 &&\left.\qquad\qquad\qquad   +s e^{-\frac{s^2}{4}}-se^{-s^2}\right]_{0}^{\infty}.\nonumber
\end{eqnarray}
 Note that  $\erf{(0)}=0$ and 

\begin{equation}\label{eq:erfasymp}
        \erf{(s)}\sim_{s\to +\infty} 1-\frac{e^{-s^2}}{s \sqrt{\pi}}+\frac{e^{-s^2}}{2\sqrt{\pi}s^3}+O\left(\frac{e^{-s^2}}{s^5}\right),
    \end{equation}
    see, for example, \cite{oldham2009atlas}.
    This means that, for large $s$, 
\begin{equation}
    \frac{s^2 \sqrt{\pi}}{2}\left(1-2\erf{(s)}+\erf{\left(\frac{s}{2}\right)} \right) = \frac{s^2 \sqrt{\pi}}{2} \left( 1- 2+\frac{2e^{-s^2}}{s\sqrt{\pi}}+1-\frac{2e^{-s^2 /4}}{s\sqrt{\pi}} +O\left(\frac{e^{-s^2}}{s^3} \right) \right),
\end{equation}
which simplifies to:
\begin{equation}
    =se^{-s^2}-se^{-s^2 /4} +O\left(\frac{e^{-s^2}}{s} \right),
\end{equation}
decaying to $0$ as $s \to \infty$.

So, 
\begin{equation}
    \rho_b=\lim_{s\rightarrow \infty}  \sqrt{\pi}\left(-\frac{\erf{(s)}}{2}+\erf{(s/2)}\right)=\frac{\sqrt{\pi}}{2}, 
\end{equation}
since $\erf{(\infty)}=1$.

\end{proof}

\subsubsection{Behaviour of large and small spacings}

First we look at the limit of very small spacings between active points. 

\begin{coro}[Local Linear Repulsion of Active Points in the Borderless Exclusion Model]
\label{coro:BorderlessLocal}
We have local linear repulsion between neighbouring active points in the borderless exclusion model. Specifically, for small $s$,
\begin{equation}
    P(s) = 2s +O(s^2).
    \label{eq:P(s)Local}
\end{equation}
\end{coro}

\begin{proof}
We start with (\ref{eq:noboundary})
   \begin{equation}
       P(s)=\frac{1}{\rho_b}\left[s\sqrt{\pi}\left(1-2\erf{(s)}+\erf{\left(\frac{s}{2}\right)}\right)-2e^{-s^2}+2e^{-\frac{s^2}{4}}\right],
   \end{equation}
   and use Lemma \ref{lem:rho} to replace $\rho_b$ with $\sqrt{\pi}/2$.
   For small $s$, the error function looks like: $\erf(s)=\frac{2}{\sqrt{\pi}}\left(s+O(s^3)\right)$ and  $e^{-s^2}=1+O(s^2).$ This means we can write:

   \begin{eqnarray}
       P(s)&=& \frac{2}{\sqrt{\pi}}\left[2(1+O(s^2))-2(1+O(s^2))+s\sqrt{\pi}\left(1-\frac{4s}{\sqrt{\pi}}+O(s^3)+\frac{2}{\sqrt{\pi}}\frac{s}{2}+O(s^3) \right)\right]
  \nonumber \\
       &=&2s+O(s^2).
   \end{eqnarray}

\end{proof}

\textrm{We started off with a $2D$ Poisson Point process where points are not correlated, and by applying a fairly natural looking exclusion process, we have produced points that are very much correlated and repel from one another. This repulsion is in fact linear in nature.}

\begin{coro}[Behaviour of Large Spacings Between Active Points in the Borderless Exclusion Model]

    Consider the Borderless Exclusion Model with at least 2 active points and their associated nearest neighbour spacing density function $P(s)$. For large $s$ we have:

\begin{equation}
    P(s) = \frac{8}{\sqrt{\pi}\;s^2}e^{-\frac{s^2}{4}}+O(e^{-\frac{s^2}{4}}s^{-4}).
   \label{eq:P(s)Tail}
\end{equation}
\label{coro:LargeExc}
\end{coro}

\begin{proof}
Start with:
    \begin{equation}
        P(s)=\frac{1}{\rho_b}\left[s\sqrt{\pi}\left(1-2\erf{(s)}+\erf{\left(\frac{s}{2}\right)}\right)-2e^{-s^2}+2e^{-\frac{s^2}{4}}\right].
    \end{equation}
    Using (\ref{eq:erfasymp}),

    \begin{eqnarray}
        &&P(s) =\nonumber\\
        && \frac{2}{\sqrt{\pi}}\left[2e^{-\frac{s^2}{4}}-2e^{-s^2}+s \sqrt{\pi}\left(\frac{2e^{-s^2}}{s\sqrt{\pi}}\left(1- \frac{1}{2s^2}\right)-\frac{2e^{-\frac{s^2}{4}}}{s\sqrt{\pi}} \left(1-\frac{2}{s^2} \right) \right)+O(e^{-\frac{s^2}{4}}s^{-4})\right] \nonumber \\
        &        &=\frac{2}{\sqrt{\pi}}\left[2e^{-\frac{s^2}{4}}-2e^{-s^2} + 2e^{-s^2}-\frac{e^{-s^2}}{s^2}-2e^{-\frac{s^2}{4}}+\frac{4e^{-\frac{s^2}{4}}}{s^2}+O(e^{-\frac{s^2}{4}}s^{-4})\right]\nonumber \\
        &&=\frac{8e^{-\frac{s^2}{4}}}{\sqrt{\pi}\;s^2}-\frac{2e^{-s^2}}{\sqrt{\pi}\;s^2}+O(e^{-\frac{s^2}{4}}s^{-4}).
    \end{eqnarray}
  So can conclude that:
    \begin{equation}
        P(s) = \frac{8}{\sqrt{\pi}\;s^2}e^{-\frac{s^2}{4}}+O(e^{-\frac{s^2}{4}}s^{-4}),
    \end{equation}
    for large $s$. 
\end{proof}
\subsection{Comparison with replication data}

\begin{figure}[H]
    \centering
    \includegraphics[scale=0.4]{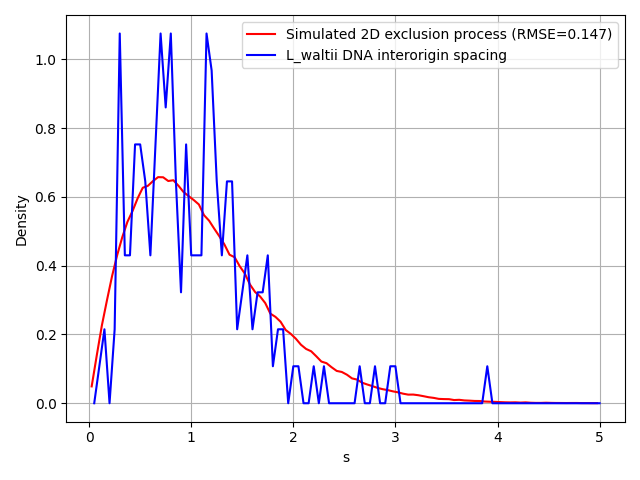} \includegraphics[scale=0.4]{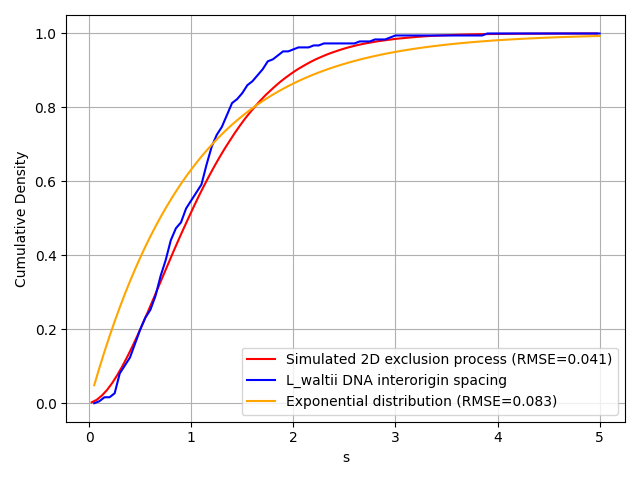}
    \caption{Left: Histogram of re-scaled spacings between midpoints of adjacent replication origins from the yeast strain Lachancea waltii (or L waltii), data taken from \cite{lwaltti2}. Right: Cumulative distribution of the same data, with the exponential distribution for comparison. RMSE is the root mean square error between the indicated curve and the L waltii data.  Total Number of Spacings: 186. Number of Chromosomes: 8.}
    \label{fig:lwatiifreq}
\end{figure}

\begin{figure}[H]
    \centering
    \includegraphics[scale=0.4]{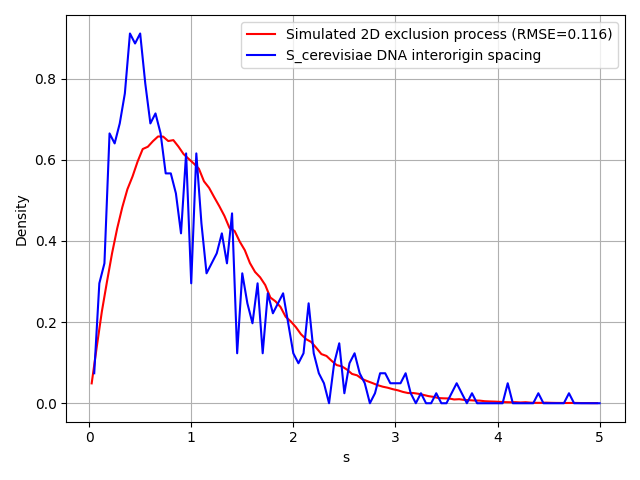}\includegraphics[scale=0.4]{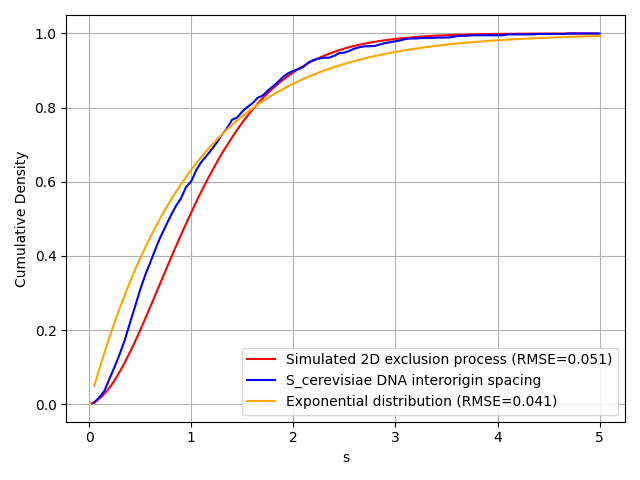}
    \caption{Left: Histogram of re-scaled spacings between midpoints of adjacent replication origins from the yeast strain Saccharomyces cerevisiae (or S cerevisiae), data taken from \cite{Scere}. Right: Cumulative distribution of the same data, with the exponential distribution for comparison. RMSE is the root mean square error between the indicated curve and the S cerevisiae data.  Total Number of Spacings: 813. Number of Chromosomes: 16.}
    \label{fig:scerefreq}
\end{figure}

\begin{figure}[H]
    \centering
    \includegraphics[scale=0.4]{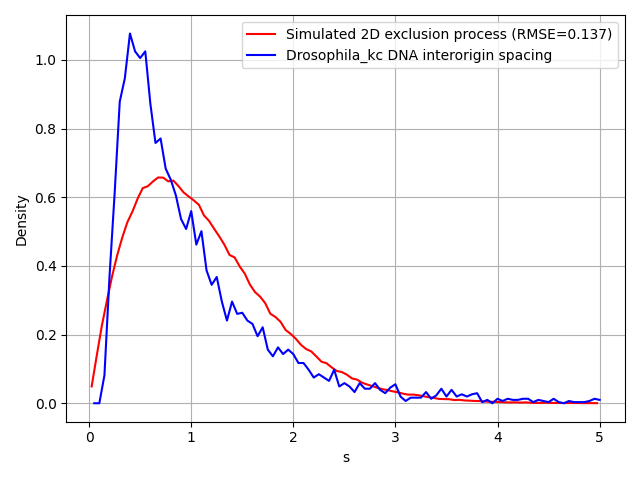}\includegraphics[scale=0.4]{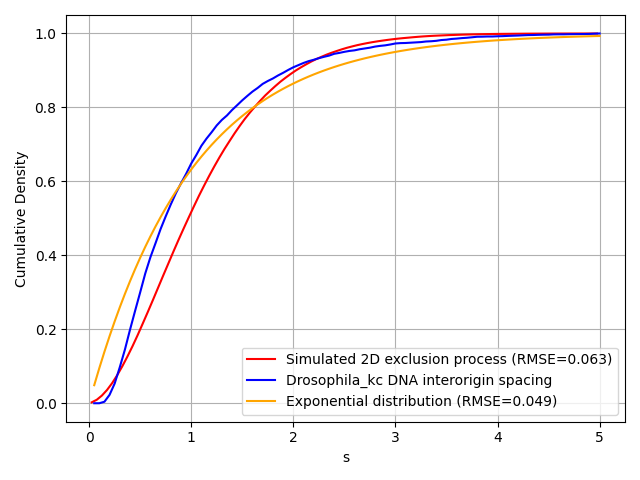}
    \caption{Left: Histogram of re-scaled spacings between midpoints of adjacent replication origins from the fruit fly Drosophila melanogaster cell line KC167 (or Drosophila KC), data taken from \cite{Drosophila}. Right: Cumulative distribution of the same data, with the exponential distribution for comparison. RMSE is the root mean square error between the indicated curve and the Drosophila KC data.   Total Number of Spacings: 6178, Number of Chromosomes: 6.}
    \label{fig:drosokcfreq}
\end{figure}

\begin{figure}[H]
    \centering
    \includegraphics[scale=0.4]{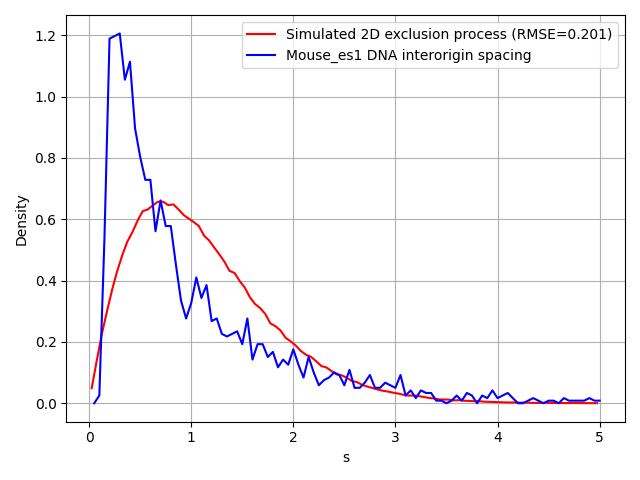}\includegraphics[scale=0.4]{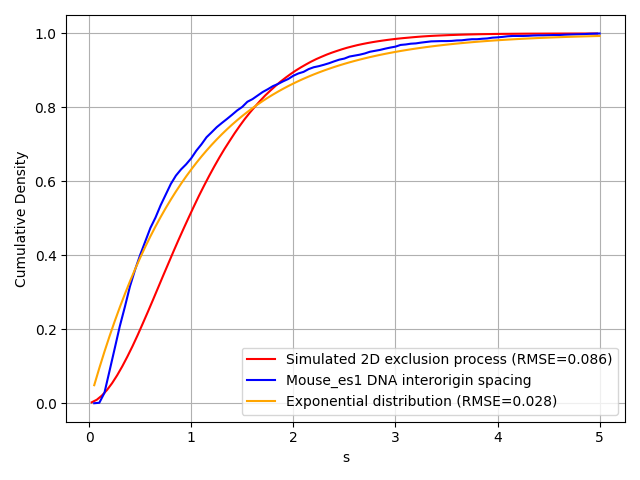}
    \caption{Left: Histogram of re-scaled spacings between midpoints of adjacent replication origins from mouse embryonic cells (or Mouse ES1), data taken from \cite{Drosophila}. Right: Cumulative distribution of the same data, with the exponential distribution for comparison. RMSE is the root mean square error between the indicated curve and the Mouse ES1 data. Total Number of Spacings: 2411. Number of Chromosomes: 1.}
    \label{fig:mousees1freq}
\end{figure}

\begin{figure}[H]
    \centering
    \includegraphics[scale=0.4]{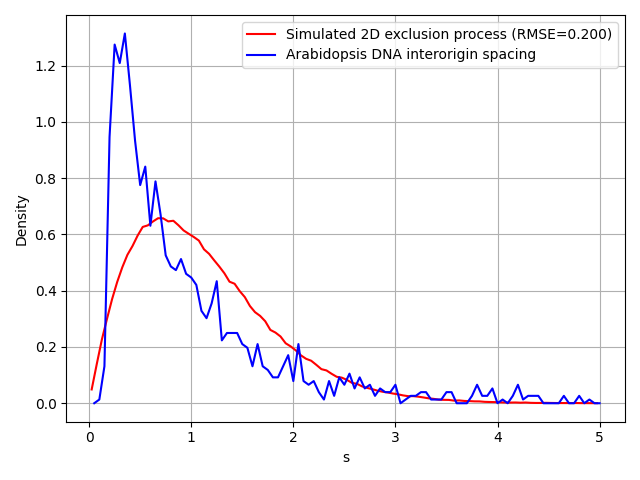} \includegraphics[scale=0.4]{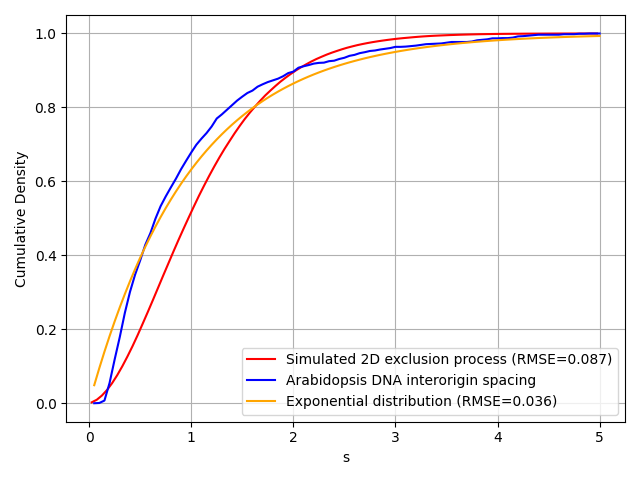}
    \caption{Left: Histogram of re-scaled spacings between midpoints of adjacent replication origins from the plant Arabidopsis thaliana, data taken from \cite{arabdopsis}. Right: Cumulative distribution of the same data, with the exponential distribution for comparison. RMSE is the root mean square error between the indicated curve and the Arabidopsis data. Total Number of Spacings: 1538. Number of Chromosomes: 5.}
    \label{fig:arabdopsisfreq}
\end{figure}

The exclusion model we propose is extremely simple and doesn't capture the intricacies of all the processes involved in DNA replication.  However, we have seen that it predicts repulsion between active replication origins and a faster then exponential decay in the large-spacing tail of the spacing distribution.  Most of the datasets of replication origins show repulsion of active replication origins (there are some exceptions, such as  a set of Drosophila data and human data, where something anomalous seems to be going on, see \cite{kn:DaySnaith} for plots of these data sets), and many (particularly the less complex organisms) show faster then exponential decay in the tail of the spacing distribution if compared with an exponential distribution $e^{-x}$ - this is seen better in the cumulative spacing distribution. In most datasets the amount of data is small, so the empirical distributions are not well-resolved, but in Figures \ref{fig:lwatiifreq} to \ref{fig:arabdopsisfreq} we see the same trend of a low probability of very small or very large spacings over data from several organisms, giving some evidence that it is robust. 

We simulated our exclusion model in Python and compared it with DNA datasets. Whilst it lacks parameterisation (and therefore ability to vary and adapt to different datasets), it captured the low probability of very small and very large inter-origin spacings seen in many available datasets. 

In order to simulate this process, we generated a 2D Poisson Point Process of intensity $\lambda = 1$ in a region of length $L=50$ and height $H=10$. Whilst in an ideal world we would set both the length and height of the domain of our point process to be as large as possible, computationally this is infeasible. We experimented with a variety of values of $L$ and $H$ and found consistent results at (and beyond) these values. In particular, due to the nature of our exclusion method, $H$ does not need to be large for our results to be stable, as active points are unlikely to be high (since this would require a large area below the point to be empty of other points). We then applied the exclusion process on these Poisson points, identifying the active points and extracting their x coordinate. We then used these to calculate their nearest neighbour spacing.

We give some examples of this behaviour in Figures \ref{fig:lwatiifreq} to \ref{fig:arabdopsisfreq}, but for more on the replication origin datasets, see the companion paper \cite{kn:DaySnaith}. 

The simulation of the exclusion model does not capture the shape of the interorigin distribution from most of the data sets, particularly the more complex organisms.  One possible modification is considering models where the likelihood of finding replication origins in a certain area of chromosome is higher in some places than in others. The exclusion model we have proposed in this paper is spatially homogeneous. Analysis of spatially inhomogeneous models is touched on in work such as \cite{chiu2000time}.

 There are other works which have proposed different timing regimes such as \cite{gaussiantimes}, which proposed using Gaussian distributed trigger times for origins on S. cerevisiae.

By modelling replication origins on chromosomes with homogeneous point processes, we have implicitly assumed that there is no location dependence. Literature has shown this to be untrue in multiple cases such as in S. cerevisiae (\cite{scere1}, \cite{scere2}, \cite{scere3}, \cite{scere4}, \cite{scere5}, \cite{scere6}) where the origin positions are well-known and mapped out, and in humans (\cite{Human1}, \cite{2021HumanSpacing}) where there is strong evidence that at least some replication origins are location dependent. In practical terms, the inhomogeneous placement of replication origins means that certain parts of the chromosome are more or less likely to have replication origins present, independent of the proximity of neighbouring origins. This mechanism was not considered in our mathematical model, which generated a homogeneous 2D Poisson point process. There are some attempts to address this in \cite{variedtimes} by having each replication origin trigger time as an independent (but with varying intensity) exponential random variables. Later attempts in the same work vary that rate throughout the evolution of the process, similar to \cite{Overcook}.

\section{Links between Exclusion Model and Polynuclear Growth Models}

In this section we introduce polynuclear growth models. These are models that represent the growth of a surface (for example a crystal surface \cite{physicsofcrystals}) over time. They have relevance to the Kolmogorov-Johnson-Mehl-Avrami (KJMA) models we introduced earlier, our exclusion model and more general Kolmogorov birth-growth models \cite{chiu1997central}.

Examining the geometry of polynuclear growth models, we will see  links with our exclusion model which motivates us to define a slightly different model - an exclusion process on a triangular domain - which allows us to calculate the expected number of active points in an exclusion system.

\subsection{Polynuclear Growth Models} \label{sect:PNGM}

Polynuclear Growth (PNG) is a random growth model in two dimensions, that evolves in continuous $1$ dimensional time through $1$ dimensional space \cite{PNG}.

Polynuclear growth has a random component that means on a small scale, it appears very rough. However, it is has a deterministic component which then smooths the process, allowing for distribution results  in large time limits.

The model consists of stochastic nucleation events, leading to an evolving surface that is described by an integer-valued height function \(h(x,t) \in \mathbf{Z}\). 

\begin{figure}[H]
    \centering
    \includegraphics{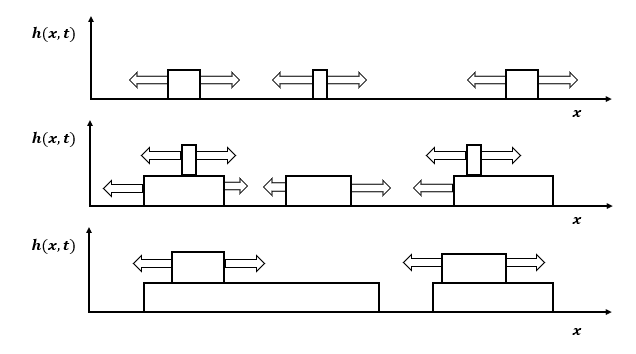}
    \caption{A visualisation of the polynuclear growth model over time with positions on the $x$-axis and the PNG height function $h(x , t)$ on the $y$-axis. Going down through the plots shows the result of time evolution ($t$ increasing). Nucleation occurs and wavefronts emanate outwards. Nucleation can occur on top of wavefronts. When two wavefronts meet, they annihilate.}
    \label{fig:PNGDynamic}
\end{figure}

We can visualise this surface as in Figure \ref{fig:PNGDynamic}. As described in the initial parts of Section 2 of \cite{PNG}, we have the position $x$ on the $x$-axis and the $y$-axis has $h(x, t)$, where each plot shows how the evolution looks for some fixed time $t$.

This is another system  where growth initiates at random points at random times and expands outwards, in a similar manner to  KJMA as well as DNA replication.  From each nucleation point, a bi-directional wavefront (akin to a replication fork) emerges outwards with unit speed. The difference here is that nucleation points can occur on any layer and the same process of forming bi-directional wavefronts occurs as it does for the bottom layer.  The height of the polynuclear growth model at $(x, t)$ is the number of wavefront layers at position $x$ after time $t$. This is illustrated in Figure \ref{fig:PNGDynamic}.

However, we can consider this model from a more geometric perspective, and plot the nucleation points in time and space, as we did for the exclusion model. 

\subsubsection{Geometric Perspective of PNG and Links with our Exclusion Model}

As described in  Section 2.3 of \cite{PNG}, we begin by generating a $2D$ Poisson point process (1 space + 1 time dimension) of some (not necessarily homogeneous) intensity $\lambda(x, t)$ in some subset of the $\mathbf{R}\times \mathbf{R}_{+}$ (i.e. $1D$ space and non-negative $1D$ time). We call these Poisson points ``nucleation points" and denote each of these points by their position and time coordinates: $(x, t)$

\begin{figure}[H]

    \includegraphics[scale=0.4]{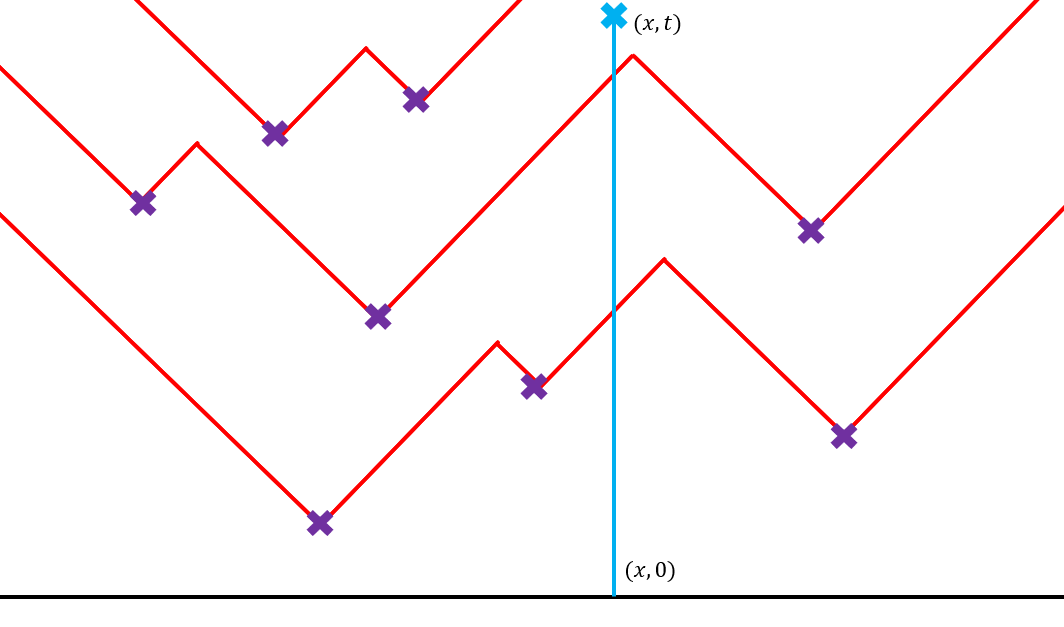}
     \begin{picture}(0,0)
\put(0,40){\scriptsize{$h=0$}}
\put(0,105){\scriptsize{$h=1$}}
\put(0,135){\scriptsize{$h=2$}}
\end{picture}
    \caption{A visualisation of geometric PNG. The spatial dimension is on the horizontal axis and time on the vertical axis.  Nucleation points are denoted by purple crosses, forward light cones denoted by red lines. The point $(x,t)$ is denoted by a blue cross and the vertical path taken to it from $(x, 0)$ is denoted by a blue line. The regions between the red boundaries represent all space-time points $(x,t)$ with the same height.}
    \label{fig:PNGHeightCount}
\end{figure}

From each nucleation point, draw the forward light cones above the point (this will be a line of slope $-1$ to the left of the nucleation point and a line of slope $+1$ to the right of the nucleation point, assuming that the wave fronts travel with unit speed). The forward light cone plots the evolution of the wavefront in time and space. Extend these straight lines until they intercept with another forward light cone from an adjacent nucleation point and then terminate the line. These intersection points represent the annihilation of two wavefronts when they meet.  The resulting image will be of piecewise continuous lines with nucleation points at each of the local minima of this boundary. These piecewise continuous lines will form mountain-range-like boundaries which cut our space into distinct regions that are stacked on top of each other, this is illustrated by the red lines in Figure \ref{fig:PNGHeightCount}.

Now consider a given point in space $(x, t)$ (not necessarily at a nucleation point). To identify  what value the PNG height function, $h(x, t)$, takes at this point, begin at $(x, 0)$ and draw a vertical line up towards $(x, t)$. As we draw this line, count how many mountain range boundaries we pass over. This process is shown in Figure \ref{fig:PNGHeightCount}. There are variations on the  PNG model, but here we assume the initial conditions \(h(x,0)=0, \forall x \in \mathbf{R}\).    Then we have the following expression:

\begin{equation}
    h(x,t)=\text{Number of boundaries crossed on the vertical path: $(x,0)\to (x,t)$}.
    \label{eq:PNGheight2}
\end{equation}

\begin{figure}[H]
    \centering
    \includegraphics[scale=0.6]{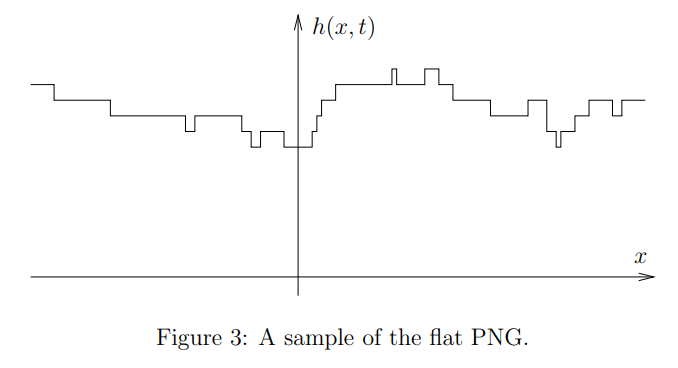}
    \caption{Original Figure taken from: \cite{PNG} with caption ``A sample of the flat PNG." This figure shows for some fixed time $t$ how the surface of a Flat PNG looks for varying values of $x$.}
    \label{fig:FlatPNG}
\end{figure}

In the above descriptions of the PNG model, if we imagine a constant density of Poisson points in the upper half plane, we have what is called the Flat PNG model (see Section 2.2 of \cite{PNG}):

\begin{define}[Flat PNG]
    Setting up a polynuclear growth model with initial conditions $h(x,0)=0, \forall x$ and constant density of Poisson points $\lambda(x,t)=\lambda>0$ in all of \(\mathbf{R}\times \mathbf{R}_{+}\), we get so-called Flat PNG geometry as illustrated in Figure \ref{fig:FlatPNG}.
    \label{def:FlatPNG}
\end{define}

Comparing our exclusion model from Section \ref{sect:model} with the Flat Polynuclear Growth model, a direct link becomes apparent.

\begin{lemma}[Correspondence between Exclusion Model and Polynuclear Growth]\label{lem:ExPNG}

Take a Poisson point process in 2D space and consider the labeling under an exclusion model structure (as in Section \ref{sect:model}) and also in the structure of a geometric view of a Flat Polynuclear Growth model with initial conditions $h(x,0)=0$.

A given point is active in the context of our exclusion model if and only if it is on the bottom layer of nucleation points in polynuclear growth.

\end{lemma}

\begin{proof}
    Consider a point on the bottom layer of nucleation points in a PNG model with initial conditions $(h(x,0)=0$. If there was a nucleation point in the backward light cone of this point then that point would occupy a new, lower nucleation layer. Therefore a point on the bottom nucleation layer is active, in the sense of the exclusion model described in Section \ref{sect:model}. The argument works both ways, since an active point will be on a layer of nucleation points. Since the active point has no points in its backward light cone, then an active point must be on the bottom layer of nucleation points.
    
  This result can also be seen by comparing a geometric visualisation of PNG (such as Figure \ref{fig:PNGHeightCount}) with any of the pictures of our exclusion model.
\end{proof}

This means that our calculations in the previous section have implications for PNG. We can now consider previous results, such as Theorem \ref{thm:NNSBorderless} relating to the nearest neighbour spacings:

\begin{theorem}[Nearest Neighbour Spacing for points in the bottom layer of Flat PNG]
\label{thm:height0}
Consider a homogeneous 2D Poisson Point process of unit intensity imposed on $\mathbb{R}\times\mathbb{R}_{+}$. We construct a Polynuclear Growth model on these and consider just the points on the boundary between height 0 and height 1. 

The nearest neighbour spacing, $P(s)ds$, of these points is the probability that, given a nucleation point on this boundary, the next nucleation point on this boundary will be between distance $s$ and $s+ds$ away.  $P(s)$ is given by:

\begin{equation}
P(s)
\label{eq:noboundary0}
    =\frac{1}{\rho_b}\left[s\sqrt{\pi}\left(1-2\erf{(s)}+\erf{\left(\frac{s}{2}\right)}\right)-2e^{-s^2}+2e^{-\frac{s^2}{4}}\right].
\end{equation}

For small $s$,
 \begin{eqnarray}
       P(s)&=&2s+O(s^2), 
   \end{eqnarray}
   and for large $s$
   \begin{equation}
    P(s) = \frac{8}{\sqrt{\pi}\;s^2}e^{-\frac{s^2}{4}}+O(e^{-\frac{s^2}{4}}s^{-4}).
\end{equation}
   The mean density of these points is   \begin{equation}
    \rho_{b}=\frac{\sqrt{\pi}}{2}.
    \end{equation}
\end{theorem}

\begin{proof} The proof follows directly from Lemma \ref{lem:ExPNG}, Theorem \ref{thm:NNSBorderless},  Lemma \ref{lem:rho}, Corollaries \ref{coro:BorderlessLocal} and \ref{coro:LargeExc}.
\end{proof}

From this theorem we see that the nucleation points on the bottom layer of Flat PNG show local linear repulsion and also that the tail of the spacing distribution decays faster than exponentially.

\subsubsection{ Flat PNG and PNG Droplet}

With no other constraints, the surface height $h(x,t)$ in the Flat PNG is translation-invariant, meaning whilst it is a random surface, the probability distribution of the $h(x, t)$ is the same as the probability distribution for $h(x',t)$ for any pairs of positions $x, x'$. The asymptotics of the height function, $h(0, t)$, have been studied for large time, and related to the GOE Tracy-Widom distribution (see \cite{prahofer2000universal}).

 An alternative model is to consider a situation where nucleation points are focused around $x=0$ and only appear for larger $x$ values at later times (see Section 2.1 of \cite{PNG}):

\begin{define}[PNG Droplet]

    Setting up a polynuclear growth model with initial conditions $h(x,0)=0$, we fix a constant density of Poisson points (say \(\lambda=1\)) in the forward light cone of the origin and zero outside of this cone. More formally:
    \begin{equation}\lambda(x,t)= 
    \begin{cases}
        1 & \text{if } |x| \leq t, \\
        0 & \text{if } |x| > t.
    \end{cases}\end{equation}
    This results in the PNG Droplet.
\end{define}

\begin{figure}[H]
    \centering
    \includegraphics[scale=0.6]{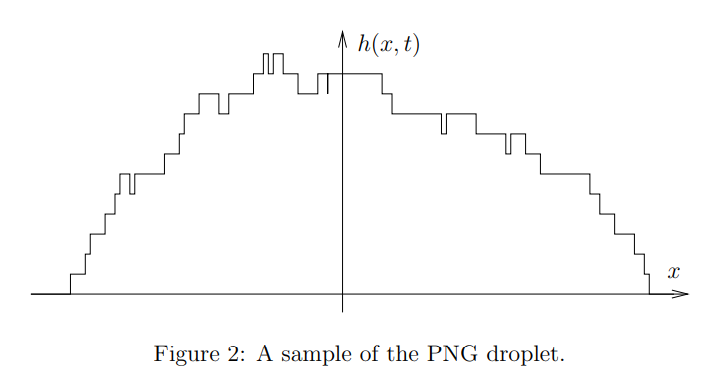}
    \caption{Original Figure taken from: \cite{PNG}. Original caption ``A sample of the PNG droplet." This figure shows for some fixed time $t$ how the surface of a PNG Droplet looks for varying values of $x$.}
    \label{fig:PNGDrop}
\end{figure}

For large time $t$ it can be shown that the PNG droplet forms a semicircle which we can see the beginnings of in Figure \ref{fig:PNGDrop}. In appropriate scaling regimes, for large $t$, the distribution of heights is related to the GUE Tracy-Widom distribution (see \cite{prahofer2000universal}).

\subsection{Links with Related Integrable Models and the Exclusion Model on a Triangle} \label{sect:triangle}

Polynuclear growth models fit into a wider group of models that all fall into the KPZ Universality class. What this essentially means is that there are equivalent functions to the height function of a given PNG model $h(x, t)$ in these other models, which have extremely similar asymptotics, for example Tracy-Widom distributions.

In this section we will introduce directed polymers on Poisson points, a model with connections to PNG as well as  to longest increasing subsequence problems \cite{ulam1961monte}. Then we will ask what comparisons can be made with our exclusion model and use this to motivate defining an exclusion model on a finite, triangular portion of the plane, making calculations more tractable.

\subsubsection{Directed polymers on Poisson points}

As described in  Section 3.1 of \cite{PNG}, we can define a partial ordering on points in \(\mathbf{R}^2\), denoted \(\prec\), characterised by up-right positioning. We say that \((a, b) \prec (c, d) \) when \(a \leq c\) and \(b \leq d\).

\textrm{We use this notion of partial ordering to define a path along Poisson points:}

\begin{define}[Directed Polymer on Poisson Points]
    Given a realisation of Poisson points from a 2D Poisson process of intensity 1 on \(\mathbf{R}^2\), a directed polymer on Poisson points is a piecewise linear path \(\gamma\) along  Poisson points, partially ordered according to \(\prec\).
\end{define}

\begin{figure}[H]
    \centering
    \includegraphics[scale=0.9]{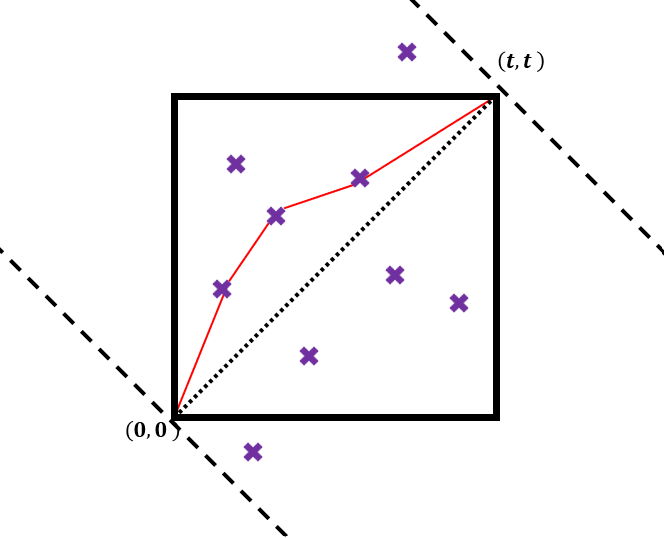}
    \caption{An illustration of directed polymers  on Poisson points. We generate a Poisson point process and consider the longest possible path going along these Poisson points but going up and to the right.
    We start at $(0, 0)$ and form a path (illustrated in red) to $(t, t)$ for some positive real value $t$.   }
    \label{fig:polymerpoints}
\end{figure}

Consider a positive parameter $t>0$ and a finite square $[0, t]\times[0,t]$ in 2 dimensional space. Scatter a $2D$ Poisson point process of uniform intensity and consider a directed polymer on these Poisson points.
    Start from $(0, 0)$ and take a partially ordered path along Poisson points which terminates at $(t,t):$
    \begin{equation}(0, 0) \prec q_{1} \prec \ldots \prec q_{l(\gamma)} \prec (t,t)\end{equation}
    where:
    \begin{equation}l(\gamma):= \text{the number of Poisson points visited by } \gamma.\end{equation}
    And write 
    \begin{equation}L(t):=\max_{\gamma \text{ path in } [0,t]^2} l(\gamma)
    \end{equation} for the longest possible partially ordered path that starts at $(0, 0)$ and terminates at $(t, t)$.

The maximum path length \(L(t)\) is a random function of $t$ and so it is of interest to consider its distribution asymptotically for large values of $t$. This is equivalent to the question of the longest increasing subsequence in a permutation of the numbers $1,2,3,\ldots,N$ (see \cite{kn:BDJ99} and the review in \cite{kn:alddia99}).

\begin{figure}[H]
    \centering
    \includegraphics[scale=0.7]{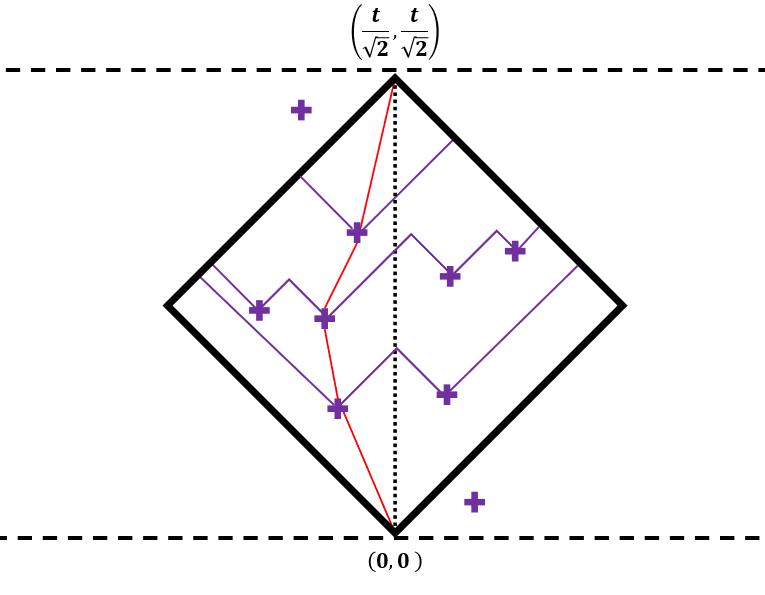}
    \caption{If we rotate Figure \ref{fig:polymerpoints} counter clockwise by 45 degrees, we see a geometric PNG structure begin to emerge. In particular, a partially ordered path on Poisson points passes through exactly one point per layer of nucleation points. Due to the rotation, the upper right corner of the square $(t, t)$ is re-scaled to $(t/\sqrt{2}, t/\sqrt{2})$, so that the vetical height is $t$. The purple lines form the boundaries of nucleation layers in geometric PNG as we saw in the previous section.}
    \label{fig:PolymerPNG}
\end{figure}

To see how this problem is related to PNG, consider Figure \ref{fig:polymerpoints} but rotate everything 45 degrees counter clockwise. Upon closer examination, it becomes evident that each Poisson point on the path under partial ordering, is on a different layer of nucleation points. In particular, consider Figure \ref{fig:PolymerPNG}: the ``up" and ``right" directions from the directed polymer picture have become the two directions of the forward light cone of each Poisson point, and Poisson points from the same boundary don't lie within the forward light cone of a given point. So we have that if $L$ is the maximum length function of directed polymers with a Poisson process of intensity $1$, then for a PNG droplet with Poisson point intensity $1$ described by height function $h$, we have that (see Section 3.1 in \cite{PNG}):
    \begin{equation}h(0,t)=L(t/\sqrt{2}).\end{equation}

\begin{figure}[H]
           \centering
           \includegraphics[scale=0.9]{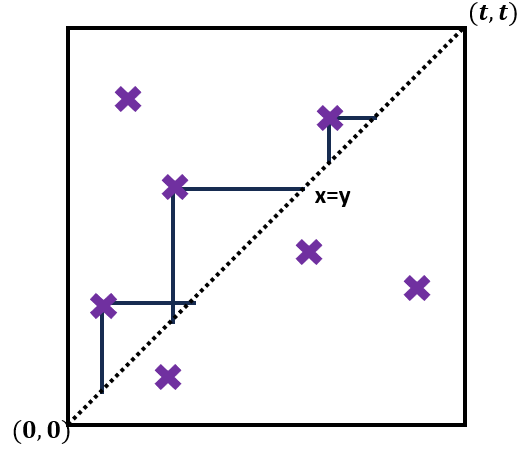}
            \caption{The directed polymer model with Poisson points marked by purple crosses. The triangles would represent backwards light cones if the system were rotated by 45 degrees clockwise, and the three of these points marked with backwards light cones would be active after this rotation, in the sense of the exclusion model in Section \ref{sect:model}, as they are empty of Poisson points.  We see that following these active points from left to right results in an up right path.  We also see that the backward light cones of any point in the upper left triangle will always be entirely contained by the upper left triangle. }
            \label{fig:PolymerP2P}
        \end{figure}

This directed polymer model inspires us to set up an exclusion model on a triangular domain.  In the directed polymer model, as we move from one Poisson point to another that is further to the right, we never move to a point that is lower in height than the original point.  This leads to a picture like Figure \ref{fig:PolymerP2P}. The triangles in Figure \ref{fig:PolymerP2P} are reminiscent of backwards light cones, if we rotated the figure clockwise by 45 degrees, and so Poisson points that have no other points contained in these triangles will be called active in the exclusion model on a triangular domain.  Essentially, this is the original exclusion model but the spatial axis is the diagonal line from $(0,0)$ to $(t,t)$ and the time axis is parallel to the $(-t,t)$ direction. 

Figure \ref{fig:PolymerP2P} helps us realise that a path through all these active points from left to right will be an up-right path, because as we proceed to the right the next active point cannot be lower than the previous one or else it would be inside the backwards light cone of that previous active point. 

The path through active points in this sense is not necessarily the longest up-right path, however, as there could be an arbitrary number of Poisson points in the forward light cone (by this we mean the area that would be the forward light cone after clockwise rotation by 45 degrees) of any active points, and deviating from the path of active points to pass through this arbitrarily large collection of points before heading to $(t,t)$ could result in a longer path. There could also be a longer path in the bottom triangle, which is not included in our exclusion model. 

For the same reason, the number of active points in this triangular domain is also less than or equal to the height at $(0,\sqrt{2}t)$ of a PNG Droplet surface on the same set of Poisson points, as we have already discussed that the longest sequence of up right points includes one from each layer of the associated PNG Droplet.

\begin{lemma} \label{lem:activePNG} 
    Given Poisson points of density 1 in the square $0\leq x\leq t$, $0\leq y\leq t$, the number of active points in the exclusion model on the triangular domain where $y\geq x$ is less than or equal to the number of points in the longest directed polymer in the square of side $t$.  It's also less than or equal to the height at $x=0$ of a PNG Droplet surface at time $\sqrt{2}t$.
\end{lemma}

\begin{proof}
    The proof proceeds by showing that each boundary in the geometric PNG set up contributes at most one active point.  Two points on the same boundary cannot both be active because  given a point, all other points on the same boundary lie in a bow-tie shaped region below the lines heading upwards to right and left at an angle of 45 degrees and above those heading downwards to right and left at an angle of 45 degrees. But when the system is rotated by 45 degrees clockwise, these sectors become the backward and forward light cones, in the sense of Figure \ref{fig:PolymerP2P}, of the specified point, and we know these cannot contain any other active points by definition. 
\end{proof}

The results in Theorem \ref{thm:height0} and Lemma \ref{lem:activePNG} establish connections with existing models, and these connections inspire the following calculation.  

\begin{theorem}[Expected Number of Active Points in an Exclusion Model on a Triangular Domain]
Consider  a Poisson point process of intensity $1$  scattered in the  triangle, $T$,  bounded by $(0, 0)$, $(0, t)$ and $(t,t)$. We define the backward light cone of a given point to be between the horizontal line extending to the right of the point and the vertical line dropping down from the point.  An active point is defined as a point with no other Poisson points in the intersection of its backwards light cone and the triangle $T$.  Write $N_{t}$ for the expected number of active points in this domain. Then we have:
\begin{equation}
           N_{t}= t \sqrt{\frac{\pi}{2}} \erf{\left( \frac{t}{\sqrt{2}} \right)}+e^{-\frac{t^2}{2}}-1.
        \end{equation}
\label{theorem:P2P}
\end{theorem}

\begin{proof}
We want to count the number of active points, that is the points with empty backward light cones.  Here we consider the line from $(0,0)$ to $(t,t)$ to be the spatial axis, so the backward light cones are those drawn in Figure \ref{fig:PolymerP2P}.  Imagine rotating the picture clockwise by 45 degrees to get back to more familiar pictures such as those in Section \ref{sect:model}.  We know that at a given point $(x, y)$, the area of the backward triangle that meets the line $y=x$ has area $\frac{(x-y)^2}{2}$. The probability of this area being empty of Poisson points is $e^{-\frac{(x-y)^2}{2}}$. Then we know that the probability of a point being in a small neighbourhood $(x, x+dx)\times (y, y+dy)$ is $dxdy$. 
        
 These two events are independent and so we have:

\begin{eqnarray}
   && P(\text{Active point in neighbourhood }(x, x+dx)\times(y, y+dy))\\
&&\qquad=P\big(\text{Poisson point in neighbourhood } (x, x+dx) \times (y, y+dx);\nonumber \\
&&\qquad\qquad\text{No Poisson points in triangle } (x,x)\to (y,y) \to (x, y)\big)\nonumber  \\
 &&\qquad=\exp{\left(-\frac{(x-y)^2}{2}\right)} dx dy=:P_{\Delta}(x,y)dxdy.\nonumber
 \label{eq:explictP}
\end{eqnarray}

Next we need to observe the relation between this quantity and the expected number of active points in triangle $T$. We  partition this probability by conditioning on the probability that there are exactly $n$ active points in the entire system.  Write
\begin{equation}
    Q_{n}:=P(\text{Exactly } n \text{ active points in triangle }T)
\end{equation}
and
\begin{equation}
    P_{n}(x, y)dxdy:=P(\text{Active point in neighbourhood }(x, x+dx)\times(y, y+dy)|\text{Exactly }n\text{ active points}).
\end{equation}
Then we have that:

\begin{equation}
    P_{\Delta}(x, y)dxdy=\sum_{n=1}^{\infty}Q_{n}\times P_{n}(x,y)dxdy.
\end{equation}
Now we consider $P_{n}(x,y)dxdy$ and ask what happens when we fix which active point is $(x,y)$. We count along from left to right and without loss of generality say that the $k-th$ active point is the $k-th$ from the left looking at the $x$-axis positions. Write:
\begin{equation}
    P^{(k)}_{n}(x,y)dxdy:=P(k\text{-th Active point at }(x, x+dx) \times(y, y+dy)| \text{Exactly }n\text{ active}).
\end{equation}

Therefore we can write:
\begin{equation}
    P_{n}(x,y)dxdy=\sum_{k=1}^{n}P^{(k)}_{n}(x,y)dxdy.
    \label{eq:IntegrateMe}
\end{equation}
Combining this all we have:
\begin{equation}
    P_{\Delta}(x,y)dxdy=\sum_{n=1}^{\infty}Q_{n}\sum_{k=1}^{n}P^{(k)}_{n}(x,y)dxdy.
    \label{eq:BigDaddy}
\end{equation}

Now we integrate (\ref{eq:IntegrateMe}) over all possible values of $x$ and $y$ in $T$. Note that:

\begin{equation}
    \int \int_{x, y \in T}P^{(k)}_{n}(x,y)dxdy =1.
\end{equation}

 This holds for every $k$, so

\begin{equation}
    \int \int_{x, y \in T}P_{\Delta}(x,y)dxdy=\sum_{n=1}^{\infty}Q_{n}\sum_{k=1}^{n} \int \int_{x, y \in T}P^{(k)}_{n}(x,y)dxdy=\sum_{n=1}^{\infty}nQ_{n}.
\end{equation}

So we can see  that integrating our function $P_{\Delta}$ over the domain $T$ gives us the expected number of active points in  $T$.

We saw above in (\ref{eq:explictP}) the form of $P_{\Delta}(x,y)$. To integrate over the domain $T$, we integrate first with respect to $x$ with the condition $0<x<y$ and then integrate with respect to $y$ with the condition $0<y<t$.
        \begin{equation}
       N_{t}=\int \int_{x, y \in T}P_{\Delta}(x, y)dx dy=\int_{0}^{t}\int_{0}^{y}\exp{\left( -\frac{(x-y)^2}{2}\right)}dxdy.
       \label{eq:integralsystem}
      \end{equation}
        
        We start by changing variables in the inner integral to recognise the error function:
    
\begin{equation}
   \int_{0}^{y}\exp{\left( -\frac{(x-y)^2}{2}\right)}dx= \sqrt{\frac{\pi}{2}} \erf{\left(\frac{y}{\sqrt{2}} \right)}.
\end{equation}

Using integration by parts, the outer integral gives

\begin{equation}
   N_{t}=  \int_{0}^{t}\sqrt{\frac{\pi}{2}} \erf{\left(\frac{y}{\sqrt{2}} \right)} dy= t \sqrt{\frac{\pi}{2}} \erf{\left( \frac{t}{\sqrt{2}} \right)}+e^{-\frac{t^2}{2}}-1.
\end{equation}

\end{proof}

We can  look at asymptotics as $t \to \infty$. Recall the asymptotic expansion \eqref{eq:erfasymp} of the error function, and apply it here:

\begin{eqnarray}
     N_{t}&=&t \sqrt{\frac{\pi}{2}} \erf{\left( \frac{t}{\sqrt{2}} \right)}+e^{-\frac{t^2}{2}}-1\nonumber \\
    &=&t \sqrt{\frac{\pi}{2}} \left(1-\frac{e^{-\frac{t^2}{2}}}{\frac{t \sqrt{\pi}}{\sqrt{2}}}+O\left(\frac{e^{-\frac{t^2}{2}}}{t^3} \right) \right)+e^{-\frac{t^2}{2}}-1
\nonumber \\
    &=& t\sqrt{\frac{\pi}{2}}-e^{-\frac{t^2}{2}}+e^{-\frac{t^2}{2}}-1+O\left(\frac{e^{-\frac{t^2}{2}}}{t^2} \right) 
\nonumber \\
    &=&t\sqrt{\frac{\pi}{2}}+O\left(1 \right).
    \label{eq:largeT}
\end{eqnarray}

This result tells us that as we let $t$ become large, the density of active points approaches a constant. In addition, the density is consistent with Section \ref{sect:model}. 

The scaling needed in order to compare this with an exclusion model of length $L$, is $L=t\sqrt{2}$. This means we can write:

\begin{equation}
    N_{\tfrac{L}{\sqrt2}}= e^{-\frac{L^2}{4}}+\frac{L\sqrt{\pi}}{2}\erf{\left(\frac{L}{2} \right)}-1
\end{equation}
and
\begin{equation}
     N_{\tfrac{L}{\sqrt2}}= L\frac{\sqrt{\pi}}{2}+O\left(1 \right),
\end{equation}
for large $L$.

In Lemma \ref{lem:rho} we calculated $\rho_{b}$, the homogeneous spatial density of active points in a borderless model and found

\begin{equation}
    \rho_{b}=\frac{\sqrt{\pi}}{2},
\end{equation}
which agrees with

\begin{equation}
    \lim_{L\to \infty}\frac{N_{\tfrac{L}{\sqrt2}}}{L}=\frac{\sqrt{\pi}}{2}.
    \label{eq:rhorelevant}
\end{equation}

This provides a nice check that our methods in Section \ref{sect:model} coincide with our approaches here in the limit of large spatial dimension (as the effect of the boundaries become less relevant).

\section*{Code}
All the code to produce simulations and plots for this work can be found on GitHub: \url{https://github.com/HuwWDay/RMTDNAData}

\section*{Acknowledgements}
Thanks to Patrick Nairne for the undergraduate summer project that initiated some of these ideas. 


Thank you to M{\'a}rton Bal{\'a}zs and Thomas Maddox for your insight and suggestions on how we could improve our stochastic model.

\section*{Funding}

\textrm{The first author was supported and funded by the EPSRC, which is part of UKRI.}


\begin{thebibliography}{DRLM{\etalchar{+}}12}

\bibitem[AD99]{kn:alddia99}
D.J. Aldous and P.~Diaconis.
\newblock Longest increasing subsequences: from patience sorting to the
  {B}aik-{D}eift-{J}ohansson theorem.
\newblock {\em Bull. Amer. Math. Soc.}, 36(4):413--32, 1999.

\bibitem[Avr41]{avrami1941}
M~Avrami.
\newblock Granulation, phase change, and microstructure kinetics of phase
  change. iii.
\newblock {\em The Journal of chemical physics}, 9(2):177--184, 1941.

\bibitem[BDJ99]{kn:BDJ99}
J.~Baik, P.~Deift, and K~Johansson.
\newblock On the distribution of the length of the longest increasing
  subsequence of random permutations.
\newblock {\em J. Amer. Math. Soc.}, 12:1119--78, 1999.

\bibitem[CCV{\etalchar{+}}11]{Drosophila}
C~Cayrou, P~Coulombe, A~Vigneron, S~Stanojcic, O~Ganier, I~Peiffer, E~Rivals,
  A~Puy, S~Laurent-Chabalier, R~Desprat, et~al.
\newblock Genome-scale analysis of metazoan replication origins reveals their
  organization in specific but flexible sites defined by conserved features.
\newblock {\em Genome Research}, 21(9):1438--1449, 2011.

\bibitem[Chi97]{chiu1997central}
SN~Chiu.
\newblock A central limit theorem for linear kolmogorov's birth-growth models.
\newblock {\em Stochastic processes and their applications}, 66(1):97--106,
  1997.

\bibitem[Chr02]{christian2002}
JW~Christian.
\newblock {\em The theory of transformations in metals and alloys}.
\newblock Newnes, 2002.

\bibitem[CMHZ{\etalchar{+}}08]{Human1}
JC~Cadoret, F~Meisch, V~Hassan-Zadeh, I~Luyten, C~Guillet, L~Duret,
  H~Quesneville, and M-N Prioleau.
\newblock Genome-wide studies highlight indirect links between human
  replication origins and gene regulation.
\newblock {\em Proceedings of the National Academy of Sciences},
  105(41):15837--15842, 2008.

\bibitem[CY00]{chiu2000time}
SN~Chiu and CC~Yin.
\newblock The time of completion of a linear birth-growth model.
\newblock {\em Advances in Applied Probability}, 32(3):620--627, 2000.

\bibitem[dMRHN10]{gaussiantimes}
A~PS de~Moura, R~Retkute, M~Hawkins, and C~A Nieduszynski.
\newblock Mathematical modelling of whole chromosome replication.
\newblock {\em Nucleic Acids Research}, 38(17):5623--5633, 2010.

\bibitem[DRLM{\etalchar{+}}12]{lwaltti2}
S~C Di~Rienzi, K~C Lindstrom, T~Mann, W~S Noble, MK~Raghuraman, and B~J Brewer.
\newblock Maintaining replication origins in the face of genomic change.
\newblock {\em Genome research}, 22(10):1940--1952, 2012.

\bibitem[DS]{kn:DaySnaith}
H.~Day and N.C. Snaith.
\newblock Thinned {COE} random matrix models for {DNA} replication.
\newblock ar{X}iv:2510.11748.

\bibitem[FCB{\etalchar{+}}06]{scere4}
W~Feng, D~Collingwood, M~E Boeck, L~A Fox, G~M Alvino, W~L Fangman, M~K
  Raghuraman, and B~J Brewer.
\newblock Genomic mapping of single-stranded {DNA} in hydroxyurea-challenged
  yeasts identifies origins of replication.
\newblock {\em Nature Cell Biology}, 8(2):148--155, 2006.

\bibitem[FP06]{PNG}
P~L Ferrari and M~Pr{\"a}hofer.
\newblock One-dimensional stochastic growth and gaussian ensembles of random
  matrices.
\newblock {\em In proceedings of "Inhomogeneous Random Systems 2005", Markov
  Processes Related Fields 12 203-234}, 2006.

\bibitem[HG10]{variedtimes}
O~Hyrien and A~Goldar.
\newblock Mathematical modelling of eukaryotic {DNA} replication.
\newblock {\em Chromosome Research}, 18(1):147--161, 2010.

\bibitem[HJBB02]{herrick2002kinetic}
J~Herrick, S~Jun, J~Bechhoefer, and A~Bensimon.
\newblock Kinetic model of {DNA} replication in eukaryotic organisms.
\newblock {\em Journal of molecular biology}, 320(4):741--750, 2002.

\bibitem[HS23]{kn:hu_stillman}
Y.~Hu and B.~Stillman.
\newblock Origins of {DNA} replication in eukaryontes.
\newblock {\em Molecular Cell}, 83(3):352--72, 2023.

\bibitem[IBA04]{irving2004bios}
W~Irving, T~Boswell, and D~Ala'Aldeen.
\newblock {\em BIOS Instant notes in Medical Microbiology}.
\newblock Taylor \& Francis, 2004.

\bibitem[JB05]{DNANuc2}
S~Jun and J~Bechhoefer.
\newblock Nucleation and growth in one dimension. ii. application to {DNA}
  replication kinetics.
\newblock {\em Physical Review E}, 71(1):011909, 2005.

\bibitem[JZB05]{DNANuc1}
S~Jun, H~Zhang, and J~Bechhoefer.
\newblock Nucleation and growth in one dimension. i. the generalized
  kolmogorov-johnson-mehl-avrami model.
\newblock {\em Physical Review E}, 71(1):011908, 2005.

\bibitem[Kol37]{kolmo1937}
AN~Kolmogoroff.
\newblock On the statistical theory of metal crystallization.
\newblock {\em Izv. Akad. Nauk SSSR, Ser. Math}, 1:335--360, 1937.

\bibitem[Lod08]{lodish2008molecular}
H~F Lodish.
\newblock {\em Molecular cell biology}.
\newblock Macmillan, 2008.

\bibitem[MR15]{SpatioTemp}
MW~Musia{\l}ek and D~Rybaczek.
\newblock Behavior of replication origins in eukaryota-spatio-temporal dynamics
  of licensing and firing.
\newblock {\em Cell Cycle}, 14(14):2251--2264, 2015.

\bibitem[NHA{\etalchar{+}}07]{Scere}
C~A Nieduszynski, S-i Hiraga, P~Ak, CJ~Benham, and AD~Donaldson.
\newblock Ori{DB}: a {DNA} replication origin database.
\newblock {\em Nucleic Acids Research}, 35(suppl\_1):D40--D46, 2007.

\bibitem[NKD06]{scere5}
C~A Nieduszynski, Y~Knox, and A~D Donaldson.
\newblock Genome-wide identification of replication origins in yeast by
  comparative genomics.
\newblock {\em Genes and Development}, 20(14):1874--1879, 2006.

\bibitem[NMNB13]{Newman}
TJ~Newman, MA~Mamun, CA~Nieduszynski, and JJ~Blow.
\newblock Replisome stall events have shaped the distribution of replication
  origins in the genomes of yeasts.
\newblock {\em Nucleic Acids Research}, 41(21):9705--9718, 2013.

\bibitem[OMS09]{oldham2009atlas}
K~B Oldham, J~Myland, and J~Spanier.
\newblock {\em An atlas of functions: with equator, the atlas function
  calculator}.
\newblock Springer, 2009.

\bibitem[PS00]{prahofer2000universal}
M~Pr{\"a}hofer and H~Spohn.
\newblock Universal distributions for growth processes in 1+ 1 dimensions and
  random matrices.
\newblock {\em Physical review letters}, 84(21):4882, 2000.

\bibitem[PV99]{physicsofcrystals}
A~Pimpinelli and J~Villain.
\newblock {\em Physics of crystal growth}.
\newblock Cambridge University Press, 1999.

\bibitem[Rhi06]{Overcook}
N~Rhind.
\newblock {DNA} replication timing: random thoughts about origin firing.
\newblock {\em Nature Cell Biology}, 8(12):1313--1316, 2006.

\bibitem[RWC{\etalchar{+}}01]{scere1}
M~K Raghuraman, E~A Winzeler, D~Collingwood, S~Hunt, L~Wodicka, A~Conway, D~J
  Lockhart, R~W Davis, B~J Brewer, and W~L Fangman.
\newblock Replication dynamics of the yeast genome.
\newblock {\em Science}, 294(5540):115--121, 2001.

\bibitem[SHMO00]{arabdopsis}
TF~Sharbel, B~Haubold, and T~Mitchell-Olds.
\newblock Genetic isolation by distance in arabidopsis thaliana: biogeography
  and postglacial colonization of europe.
\newblock {\em Molecular Ecology}, 9(12):2109--2118, 2000.

\bibitem[SHR19]{genetherapy}
D~HM Steffin, E~M Hsieh, and R~H Rouce.
\newblock Gene therapy: current applications and future possibilities.
\newblock {\em Advances in Pediatrics}, 66:37--54, 2019.

\bibitem[Ula61]{ulam1961monte}
S~M Ulam.
\newblock Monte carlo calculations in problems of mathematical physics.
\newblock {\em Modern Mathematics for the Engineers}, pages 261--281, 1961.

\bibitem[WAC{\etalchar{+}}01]{scere2}
J~J Wyrick, J~G Aparicio, T~Chen, J~D Barnett, E~G Jennings, R~A Young, S~P
  Bell, and O~M Aparicio.
\newblock Genome-wide distribution of {ORC} and {MCM} proteins in s.
  cerevisiae: high-resolution mapping of replication origins.
\newblock {\em Science}, 294(5550):2357--2360, 2001.

\bibitem[WKP{\etalchar{+}}21]{2021HumanSpacing}
W~Wang, KN~Klein, K~Proesmans, H~Yang, C~Marchal, X~Zhu, T~Borrman, A~Hastie,
  Z~Weng, J~Bechhoefer, et~al.
\newblock Genome-wide mapping of human {DNA} replication by optical replication
  mapping supports a stochastic model of eukaryotic replication.
\newblock {\em Molecular Cell}, 81(14):2975--2988, 2021.

\bibitem[WM39]{william1939}
J~William and R~Mehl.
\newblock Reaction kinetics in processes of nucleation and growth.
\newblock {\em Trans. Metall. Soc. AIME}, 135:416--442, 1939.

\bibitem[XAAT06]{scere6}
W~Xu, J~G Aparicio, O~M Aparicio, and S~Tavar{\'e}.
\newblock Genome-wide mapping of orc and mcm2p binding sites on tiling arrays
  and identification of essential ars consensus sequences in s. cerevisiae.
\newblock {\em BMC Genomics}, 7(1):1--16, 2006.

\bibitem[YN88]{yang1988}
CP~Yang and JF~Nagle.
\newblock Phase transformations in lipids follow classical kinetics with small
  fractional dimensionalities.
\newblock {\em Physical Review A}, 37(10):3993, 1988.

\bibitem[YTK02]{scere3}
N~Yabuki, H~Terashima, and K~Kitada.
\newblock Mapping of early firing origins on a replication profile of budding
  yeast.
\newblock {\em Genes to Cells}, 7(8):781--789, 2002.

\end{thebibliography}

\newcommand{\etalchar}[1]{$^{#1}$}

\end{document}